\documentclass[pra,aps,twocolumn,showpacs,preprintnumbers,amsmath,amssymb]{revtex4}

\usepackage{graphicx}
\usepackage{dcolumn}
\usepackage{bm}

\newcommand{\ket}[1]{\vert #1 \rangle}
\newcommand{\bra}[1]{\langle #1 \vert}

\newcommand{\meanvalue}[3]{\langle #1 \vert #2 \vert #3 \rangle}
\newcommand{\ketbra}[2]{\vert #1 \rangle \langle #2 \vert}

\begin{document}


\title{Monitoring atom-atom entanglement and decoherence\\ in a solvable tripartite open system in cavity QED}

\author{Matteo Bina, Federico Casagrande\footnote{federico.casagrande@mi.infn.it}, Alfredo Lulli}
 \affiliation{Dipartimento di Fisica, Universit\`a degli Studi di Milano, Via Celoria 16, 20133 Milano, Italy}

\author{Enrique Solano}
\affiliation{Physics Department, ASC, and CeNS,
Ludwig-Maximilians-Universit\"at, Theresienstrasse 37, 80333 Munich,
Germany \\ Secci\'on F\'{\i}sica, Departamento de
Ciencias, Pontificia Universidad Cat\'olica del Per\'u, Apartado
1761, Lima, Peru}

\date{\today}

\begin{abstract}

We present a fully analytical solution of the dynamics of two
strongly-driven atoms resonantly coupled to a dissipative cavity
field mode. We show that an initial atom-atom entanglement cannot be
increased. In fact, the atomic Hilbert space divides into two
subspaces, one of which is decoherence free so that the initial
atomic entanglement remains available for applications, even in
presence of a low enough atomic decay rate. In the other subspace a
measure of entanglement, decoherence, and also purity, are described
by a similar functional behavior that can be monitored by joint
atomic measurements. Furthermore, we show the possible generation of
Schr\"odinger-cat-like states for the whole system in the transient
regime, as well as of entanglement for the cavity field and the
atom-atom subsystems conditioned by measurements on the
complementary subsystem.

\end{abstract}


\maketitle

\section{\label{sec:1}Introduction}

The interaction of a two-level system with a quantized single mode
of a harmonic oscillator, named Jaynes-Cummings (JC)
model~\cite{JaynesCummings}, is arguably the most fundamental
quantum system describing the interaction of matter and light. The
JC model has found its natural playground in the field of cavity
quantum electrodynamics (CQED), in the
microwave~\cite{ReviewParis,ReviewGarching} and in the optical
regime~\cite{ReviewOptical}, as well as in other physical systems,
like trapped ions~\cite{ReviewIons} or Circuit~\cite{ReviewYale} and
solid-state~\cite{SSCQED} QED. Extensions of the JC model to more
atoms and more modes, externally driven or not, have been developed
and, presently, we enjoy a vast number of theoretical and
experimental developments. Unfortunately, a great part of them are
not easy to handle and most of the interesting physics has to be
extracted from heavy numerical solutions and calculations,
especially when realistic dissipative processes are taking into
account.\\
The advent of quantum information~\cite{NielsenChuang} has been a
fresh input in the field of CQED~\cite{Haroche-Raimond}, reshaping
concepts and using it for fundamental tests and initial steps in the
demanding field of quantum information processing. Here, coherence
and the generation of entanglement play an important role, and in
particular the manner in which they are affected by the presence of
a dissipative environment~\cite{SchleichBook}. In spite of their
relevance, most of the attractive quantum models do not enjoy
analytical solutions and the few available ones are not as close to
realistic conditions as desired.\\
In this paper, we consider a system composed by two coherently
driven two-level atoms trapped inside a cavity and coupled to one of
its quantized modes. The experimental implementation seems to be
feasible due to the recent advances in deterministic trapping of
atoms in optical cavities ~\cite{trapped-atoms,Fortier}. We show
that under full resonance conditions and negligible atomic decays
the system dynamics can be solved analytically also in the presence
of cavity field dissipation. With the solutions at hand we are able
to monitor the purity of the cavity field and atom-atom subsystems,
as well as the entanglement and decoherence dynamics of the latter.
Atomic entanglement cannot be generated. However, the quantum
correlations of suitable entangled states can remain completely
protected in a decoherence-free subspace. In case of atomic states
outside this subspace, the decays of quantum entanglement, coherence
and purity are remarkably described by the same function.\\
In Sec.~\ref{sec:2} we introduce the master equation of the
considered open system, in Sec.~\ref{sec:3} we present the
analytical solutions of the system evolution, in Sec.~\ref{sec:4} we
study the dynamics of entanglement in the atom-atom subsystem, in
Sec.~\ref{sec:5} we consider the conditional generation of subsystem
states, in Sec.\ref{sec:6} we present fully numerical results that
confirm our analytical developments, and in Sec.~\ref{sec:7} we
conclude with a summary of our results and physical discussions. In
the Appendix we discuss the solution of the master equation
presented in Sec.~\ref{sec:2}.

\section{Open System Master Equation}\label{sec:2}

\begin{figure}
\centering
\resizebox{0.95\columnwidth}{!}{\includegraphics{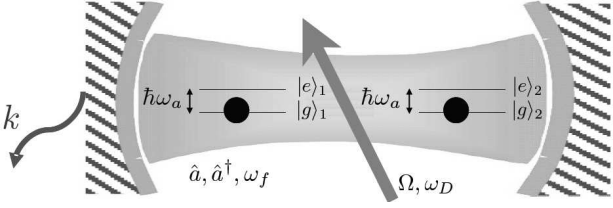} }
\caption{Two driven two-level atoms interacting with a dissipative
cavity mode.} \label{fig:cavita}
\end{figure}

We consider a pair of two-level atoms  interacting inside a cavity
with a field mode also coupled to the environment
(see~Fig.~\ref{fig:cavita}). A coherent external field of frequency
$\omega_D$ drives both atoms during the interaction with the cavity
mode of frequency $\omega_f$~\cite{Solano-Agarwal,PavelOptical}. The
transition frequency $\omega_a$ between excited and ground states,
$\ket{e}_j$ and $\ket{g}_j$ ($j=1,2$), respectively, is the same for
the two-level atoms. This system could be experimentally implemented
with two-level Rydberg atoms in a microwave cavity
\cite{Haroche-Raimond} or with three-level atoms reduced effectively
to two levels interacting with an optical cavity. Relevant advances
in cooling and trapping atoms in optical cavities have been recently
achieved \cite{trapped-atoms},\cite{Fortier}. Similar dynamics could
be also implemented in trapped ions interacting with a vibrational
mode instead of the
cavity mode \cite{Haroche-Raimond}.\\
\indent The Hamiltonian which describes the whole system unitary
dynamics is
\begin{eqnarray}\label{eq:hamiltonianSDJC}
\hat{\mathcal{H}}(t)&=&\frac{\hbar\omega_a}{2}\sum_{j=1}^2\hat{\sigma}^z_{j}
+\hbar\omega_f \hat{a}^{\dag}\hat{a}+\hbar\Omega\sum_{j=1}^2\bigg (
e^{-i\omega_D t}\hat{\sigma}_j^{\dag}+\,\nonumber\\
&+&e^{i\omega_D t}\hat{\sigma}_j\bigg )+\hbar g\sum_{j=1}^2\bigg (
\hat{\sigma}_j^{\dag}\hat{a}+\hat{\sigma}_j\hat{a}^{\dag}\bigg ) ,
\end{eqnarray}
where $\Omega$ is the Rabi frequency associated with the coherent
driving field amplitude, $g$ the atom-cavity mode coupling constant
(taken equal for both atoms),  $\hat{a}$ ($\hat{a}^{\dag}$) the
field annihilation (creation) operator,
$\hat{\sigma}_j=\ket{g}_j\bra{e}$
($\hat{\sigma}_j^{\dag}=\ket{e}_j\bra{g}$) the atomic lowering
(raising) operator, and
$\hat{\sigma}^z_{j}=\ket{e}_j\bra{e}-\ket{g}_j\bra{g}$ the inversion
operator.\\
\noindent In the perspective of possible experimental implementation
of our scheme we must include the effects of cavity mode dissipation
and the decay of the atomic upper level. Therefore, we must solve
the following master equation (ME) for the statistical density
operator $\hat{\rho}'$ of the whole tripartite system
\begin{equation}\label{eq:ME}
\dot{\hat{\rho}}'=-\frac{i}{\hbar}[\hat{\mathcal{H}},\hat{\rho}'] +
\hat{\mathcal{L}}_f\hat{\rho}'+\hat{\mathcal{L}}_a\hat{\rho}'
\end{equation}
where
\begin{eqnarray}
\hat{\mathcal{L}}_f\hat{\rho}'&=&\frac{k}{2}[2\hat{a}\hat{\rho}'
\hat{a}^{\dag}-\hat{a}^{\dag}\hat{a}\hat{\rho}' -
\hat{\rho}' \hat{a}^{\dag}\hat{a}] \,\nonumber\\
\hat{\mathcal{L}}_a\hat{\rho}'&=&\frac{\gamma}{2}\sum_{j=1}^2[2\hat{\sigma}_j\hat{\rho}'
\hat{\sigma}_j^{\dag}-
\hat{\sigma}_j^{\dag}\hat{\sigma}_j\hat{\rho}' - \hat{\rho}'
\hat{\sigma}_j^{\dag}\hat{\sigma}_j] \label{Liouville_0},
\end{eqnarray}
where $k$ and $\gamma$ are the cavity and atomic decay rates,
respectively, and the environment is modeled by a thermal bath at
zero temperature.\\ Changing to the interaction picture the
dissipative terms remain unchanged and the ME (\ref{eq:ME}) can be
rewritten as
\begin{equation}\label{eq:MEIP}
\dot{\hat{\rho}}_I=-\frac{i}{\hbar}[\hat{\mathcal{H}}_I,\hat{\rho}_I]
+ \hat{\mathcal{L}}_f\hat{\rho}_I+\hat{\mathcal{L}}_a\hat{\rho}_I
\end{equation}
where Hamiltonian (\ref{eq:hamiltonianSDJC}) has been replaced by
the time-independent Hamiltonian
$\hat{\mathcal{H}}_I=\hat{\mathcal{H}}_0+\hat{\mathcal{H}}_1$ with
\begin{eqnarray}\label{eq:IP_H}
\hat{\mathcal{H}}_0&=&-\hbar\delta \hat{a}^{\dag}\hat{a}+
\hbar\Omega\sum_{j=1}^2\left ( \hat{\sigma}_j^{\dag}+
\hat{\sigma}_j\right )\,\nonumber\\
\hat{\mathcal{H}}_{1}&=&\hbar g\sum_{j=1}^2\left (
\hat{\sigma}_j^{\dag}\hat{a}+\hat{\sigma}_j\hat{a}^{\dag}\right ) ,
\end{eqnarray}
where we introduced the atom-cavity field detuning parameter
$\delta=\omega_a - \omega_f$, and from now on we consider a
resonance condition between the atoms and the external field
($\omega_D=\omega_a$). We remark that Hamiltonian (\ref{eq:IP_H})
was derived by a standard technique, whereas in the case of less
simple systems, e.g. three-level atoms, more refined treatments are
necessary to derive a time-independent Hamiltonian such as adiabatic
elimination
or nonlinear rotations, as e.g. in \cite{SDOAL}.\\
The ME (\ref{eq:MEIP}) can be solved only by numerical techniques,
as we discuss in Sec.\ref{sec:6} where we present some numerical
results for the whole system dynamics. On the other hand, if we
consider negligible atomic decay ($\gamma=0$) it is possible to
solve Eq.(\ref{eq:MEIP}) analytically. First of all, we consider the
unitary transformation $\hat{\mathcal{U}}(t)=exp
\{\frac{i}{\hbar}\hat{\mathcal{H}}_0 t \}$ and we derive for the
density operator
$\hat{\rho}(t)=\hat{\mathcal{U}}(t)\hat{\rho}_I(t)\hat{\mathcal{U}}^{\dagger}(t)$
the following ME:
\begin{equation}\label{eq:ME2}
\dot{\hat{\rho}}=-\frac{i}{\hbar}[\hat{\mathcal{U}}\hat{\mathcal{H}}_1\hat{\mathcal{U}}^{\dagger},\hat{\rho}]
+ \hat{\mathcal{L}}_f\hat{\rho}
\end{equation}
where the transformed Hamiltonian can be written as
\begin{eqnarray}
\hat{\mathcal{U}}\hat{\mathcal{H}}_1\hat{\mathcal{U}}^{\dagger}&=&\frac{\hbar
g}{2}\sum_{j=1}^2\big
[ \ket{+}_j\bra{+}-\ket{-}_j\bra{-}+e^{2i\Omega t}\ket{+}_j\bra{-} \,\nonumber\\
&-&e^{-2i\Omega t}\ket{-}_j\bra{+}\big ]ae^{-i\delta t} + {\rm H.c.} .
\end{eqnarray}
Here, $\{\ket{+}_j,\ket{-}_j\}$ ($j=1,2$) is a rotated basis
connected to the standard basis $\{\ket{e}_j,\ket{g}_j\}$ via
$\ket{\pm}_j=\frac{\ket{g}_j\pm\ket{e}_j}{\sqrt{2}}$. In the
strong-driving regime for the interaction between the atoms and
the external coherent field, $\Omega\gg g$, we can use the
rotating-wave approximation (RWA) obtaining the effective
Hamiltonian \cite{Solano-Agarwal,PavelOptical}
\begin{equation}\label{eq:effective_H}
\hat{\mathcal{H}}_{\rm eff}(t)=\frac{\hbar
g}{2}\sum_{j=1}^2(\hat{\sigma}_j^{\dag}
+\hat{\sigma}_j)(\hat{a}e^{i\delta t}+\hat{a}^{\dag}e^{-i\delta t}) .
\end{equation}
Equation (\ref{eq:effective_H}) outlines the presence of
Jaynes-Cummings ($\hat{\sigma}_j^{\dag}\hat{a}
+\hat{\sigma}_j\hat{a}^{\dag}$) as well as anti-Jaynes-Cummings
($\hat{\sigma}_j^{\dag}\hat{a}^{\dag} +\hat{\sigma}_j\hat{a}$)
interaction terms of each coherently driven atom with the cavity
mode. We remark that in the framework of trapped ions a similar
Hamiltonian can be found but the cavity mode is replaced by the
vibrational mode of the ions system \cite{KikeIons}. In the
following we will describe the solution of the effective ME:
\begin{equation}\label{eq:ME3}
\dot{\hat{\rho}}=-\frac{i}{\hbar}[\hat{\mathcal{H}}_{\rm
eff}(t),\hat{\rho}] + \hat{\mathcal{L}}_f\hat{\rho}.
\end{equation}

\section{Exact Solution at resonance for atoms prepared in the ground state}\label{sec:3}
\begin{figure}[h]
\resizebox{0.60\columnwidth}{!}{ \includegraphics{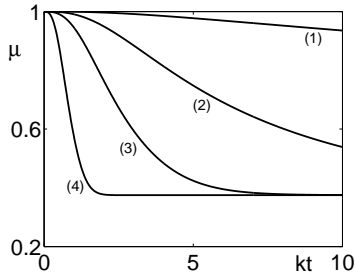} }\\
\caption{Purity $\mu(t)$ of the whole system density operator vs.
dimensionless time $kt$ for dimensionless coupling constant $g/k$:
(1) $0.05$, (2) $0.2$, (3) $0.5$, (4) $2$.} \label{fig2}
\end{figure}

In this section we consider the solution of the ME (\ref{eq:ME3}) in
the case of exact resonance $\delta=0$, cavity field initially in
the vacuum state, and both atoms prepared in the ground state as an
example of separable initial state. We leave to future treatments
the cases of cavity field prepared in more general states. The exact
solution of the ME for any atomic preparation is described in
details in the Appendix. It is based on the following decomposition
for the density operator $\hat{\rho}(t)$ of the whole system:
\begin{equation}\label{decomposition}
\hat{\rho} (t)=\sum_{i,j=1}^4 \meanvalue{i}{\hat{\rho}(t)}{j}
\ketbra{i}{j} = \sum_{i,j=1}^4\hat{\rho}_{ij}(t)\ketbra{i}{j}
\end{equation}
where $\{\ket{i}\}_{i=1}^4=\{ \ket{++},\ket{+-},\ket{-+},\ket{--}
\}$ is the rotated basis of the atomic Hilbert space. Here we report
only the final expressions for the field operators $\hat{\rho}_{ij}$
in the present case:
\begin{eqnarray}
\label{rhoijground}
\hat{\rho}_{(11,44)}(t) & = & \frac{1}{4}\ketbra{\mp\alpha(t)}{\mp\alpha(t)} \,\nonumber \\
\hat{\rho}_{(12,13)}(t) & = & \frac{1}{4}\frac{f_1(t)}{e^{- |\alpha(t)|^2 / 2}}\ketbra{-\alpha(t)}{0} \,\nonumber \\
\hat{\rho}_{14}(t) & = & \frac{1}{4}\frac{f_2(t)}{e^{-2|\alpha(t)|^2}}\ketbra{-\alpha(t)}{\alpha(t)} \,\nonumber \\
\hat{\rho}_{(22,23,33)}(t) & = & \frac{1}{4}\ketbra{0}{0} \,\nonumber \\
\hat{\rho}_{(24,34)}(t) & = &
\frac{1}{4}\frac{f_1(t)}{e^{-|\alpha(t)|^2 /
2}}\ketbra{0}{\alpha(t)} ,
\end{eqnarray}
where we introduced the time dependent coherent field amplitude
\begin{equation}\label{alfat}
\alpha(t) = i\frac{2g}{k}\left ( 1-e^{-\frac{k}{2}t}\right ) ,
\end{equation}
and the function
\begin{equation}
f_1(t) = \exp\left \{ -\frac{2g^2}{k}t+\frac{4g^2}{k^2}\left
( 1-e^{-\frac{k}{2}t}\right ) \right \}.\label{eq:f1}\\
\end{equation}
We recall that $\hat{\rho}_{ji}(t)=\hat{\rho}_{ij}^{\dagger}(t)$. We
notice the presence of single atom-cavity field coherences whose
evolution is ruled by the function $f_1(t)$ \cite{SDOAL}, as well as
full atom-atom-field coherences ruled by $f_2(t)=f_1^4(t)$, that we
shall discuss later on. There are also two one-atom coherences, and
two diagonal terms, which do not evolve in time, corresponding to a
pure state
\begin{equation}\label{gg0}
|0\rangle\otimes(|+-\rangle + |-+\rangle) =
|0\rangle\otimes(|gg\rangle - |ee\rangle),
\end{equation}
where we recognize (up to normalization) a Bell atomic state
$|\Phi^{-}\rangle$. The explanation goes as follows. First of all,
if we start with atoms prepared in states of the rotated basis used
in the decomposition of (\ref{decomposition}), we obtain much
simpler results due to the structure of Hamiltonian
(\ref{eq:effective_H}) on resonance and the obvious uninfluence of
dissipation on the cavity vacuum. Actually, either the field states
are coherent, $|0\rangle \otimes|\pm\pm\rangle \mapsto
|\mp\alpha(t)\rangle\otimes|\pm\pm\rangle $, or there is no
evolution at all for the states $|0\rangle\otimes|\pm\mp\rangle$,
showing the presence of an invariant subspace for system dynamics.
It is the component of the initial state
$|0\rangle\otimes|gg\rangle$ in that subspace, (\ref{gg0}), which
does not evolve. We note that starting from atoms prepared in the
other elements of the standard basis we obtain solutions quite
analogous to (\ref{rhoijground}), as can be easily
verified from the Appendix.\\
\indent In the transient regime ($kt\ll 1$), the decoherence
function can be approximated by $f_1(t)\simeq
e^{-\frac{|\tilde{\alpha}(t)|^2}{2}}$ where $\tilde{\alpha}(t)=igt$,
and the whole system is described by a pure Schr\"odinger-cat-like
state
\begin{eqnarray}\label{gatto_sistema_2}
\ket{\tilde{\psi}(t)} &=&\frac{1}{2}\bigg [
\ket{-\tilde{\alpha}(t)}\otimes\ket{++}+\ket{0}\otimes(\ket{+-}+\ket{-+})\,\nonumber\\
&+&\ket{\tilde{\alpha}(t)}\otimes\ket{--}\bigg ].
\end{eqnarray}
If we rewrite this result in the standard atomic basis
\begin{eqnarray}\label{gatto_sistema_2bis}
\ket{\tilde{\psi}(t)}=\frac{1}{4}\bigg[\big(\ket{-\tilde{\alpha}(t)}-2\ket{0}+\ket{\tilde{\alpha}(t)}\big)
\otimes\ket{ee}\,\nonumber\\
+\big(\ket{-\tilde{\alpha}(t)}-\ket{\tilde{\alpha}(t)}\big)
\otimes(\ket{eg}+\ket{ge})\,\nonumber\\
+\big(\ket{-\tilde{\alpha}(t)}+2\ket{0}+\ket{\tilde{\alpha}(t)}\big)\otimes\ket{gg}\bigg]
\end{eqnarray}
showing the onset of correlations between atomic states and cavity
field cat-like states. For the terms related to the field subsystem
we recover expressions analogous to those derived in \cite{SDM} for
a strongly driven micromaser system. Unlike the present system, in
that case the atoms pump the cavity mode with a Poissonian
statistics, interacting for a very short time such that the cavity
dissipation is relevant only in the time intervals between atomic
injections. Furthermore, the cavity field states are conditioned on
atomic measurements.\\
At steady state ($kt \to \infty$) the density operator is mixed and
given by:
\begin{eqnarray}\label{miscela_sistema_2}
\hat{\rho}^{SS}&=&\frac{1}{4}\bigg [ \ketbra{-\alpha^{SS}}{-\alpha^{SS}}\otimes\ketbra{++}{++}+\ketbra{\alpha^{SS}}{\alpha^{SS}}\,\nonumber\\
&\otimes&\ketbra{--}{--}+\ketbra{0}{0}\otimes\bigg ( \ketbra{+-}{+-}+\,\nonumber\\
&+&\ketbra{+-}{-+}+\ketbra{-+}{+-}+\ketbra{-+}{-+}\bigg ) \bigg ]\,\nonumber\\
\end{eqnarray}
with $\alpha^{SS}=2i\frac{g}{k}$. Interestingly, the steady state
has not a fully diagonal structure, i.e., it is not completely
mixed, in agreement with the previous discussion on the
time-dependent solution. The change from a pure state to a mixed one
and the degree of mixedness can be we evaluated by the purity of the
whole system $\mu(t)=Tr[\hat{\rho}^2(t)]$:
\begin{equation}\label{purity}
\mu(t)=\frac{1}{8}\left[3+4\frac{f_1^2(t)}{e^{-|\alpha(t)|^2}}+\frac{f_2^2(t)}{e^{-4|\alpha(t)|^2}}\right].
\end{equation}
In Fig.~\ref{fig2} we show $\mu(t)$ as a function of the
dimensionless time $kt$ for different values of the dimensionless
coupling constant $g/k$. We see that the purity decays rather fast
in the strong coupling regime $g/k\gtrsim1$.

\subsection{Subsystems dynamics}\label{subsec:3-1}
\begin{figure}[h]
\resizebox{0.60\columnwidth}{!}{ \includegraphics{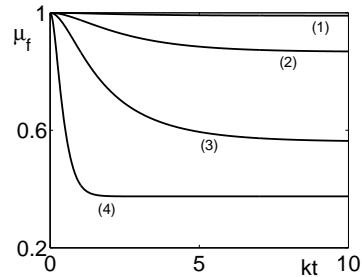} }\\
\caption{Purity $\mu_f(t)$ of the cavity field subsystem vs.
dimensionless time $kt$ for dimensionless coupling constant $g/k$:
(1) $0.05$, (2) $0.2$, (3) $0.5$, (4) $2$.} \label{fig3}
\end{figure}

Now we consider the time evolution of the atomic and cavity field
subsystems in the case of both atoms initially prepared in the
ground state. The cavity field reduced density operator
$\hat{\rho}_f(t)=Tr_{a}[\hat{\rho}(t)]$ can be derived by tracing
over both atoms:
\begin{eqnarray}\label{rho_f}
\hat{\rho}_f(t)&=&\frac{1}{4}\bigg [
\ketbra{-\alpha(t)}{-\alpha(t)}+2\ketbra{0}{0}+\ketbra{\alpha(t)}{\alpha(t)}\bigg
] . \nonumber\\
\end{eqnarray}
Hence the cavity field mean photon number is
$\langle\hat{N}\rangle=\frac{1}{2}|\alpha(t)|^2$ and its steady
state value $\langle\hat{N}\rangle_{SS}=2\frac{g^2}{k^2}$. These
results hold for atoms prepared in any state of the standard basis.
At any time Eq. (\ref{rho_f}) describes a mixed state, whose purity
$\mu_f(t)=Tr_f[ \hat{\rho}_f^2(t)]$ is:
\begin{equation}\label{purity_F}
\mu_f(t)=\frac{1}{8}\left[3+4e^{-|\alpha(t)|^2}+e^{-4|\alpha(t)|^2}\right].
\end{equation}
In Fig.~\ref{fig3} we show the purity $\mu_f(t)$ as a function of
dimensionless time $kt$ for different values of the ratio $g/k$,
showing a better survival of the purity than for the global state of
Fig.~\ref{fig2} except
in the strong coupling regime.\\
\indent The reduced atom-atom density operator $\hat{\rho}_{a}(t)$
can be obtained by tracing over the field variables. In the rotated
basis $\{ \ket{++},\ket{+-},\ket{-+},\ket{--} \}$, we obtain
\begin{equation}
\label{mx:gg_±}
\rho_{a}^{\pm}(t)=\frac{1}{4}\left(\begin{array}{cccc} 1 & f_1(t) & f_1(t) & f_2(t)\\
f_1(t) & 1 & 1 & f_1(t)\\
f_1(t) & 1 & 1 & f_1(t)\\
f_2(t) & f_1(t) & f_1(t) & 1
\end{array}\right).
\end{equation}
The presence of six time-independent matrix elements is in agreement
with the remarks below Eq.(\ref{gg0}). The purity $\mu_a(t)=Tr_{a}[
\hat{\rho}_{a}^2(t)]$ of the bi-atomic subsystem is:
\begin{equation}\label{purity_a}
\mu_a(t)=\frac{1}{8}\left[3+4f_1^2(t)+f_2^2(t)\right].
\end{equation}

\begin{figure}[h]
\resizebox{0.55\columnwidth}{!}{ \includegraphics{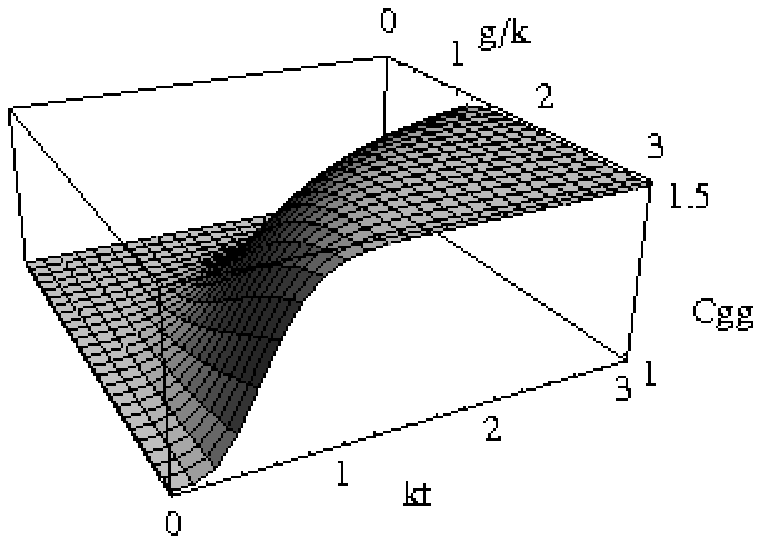} } a)
\resizebox{0.55\columnwidth}{!}{ \includegraphics{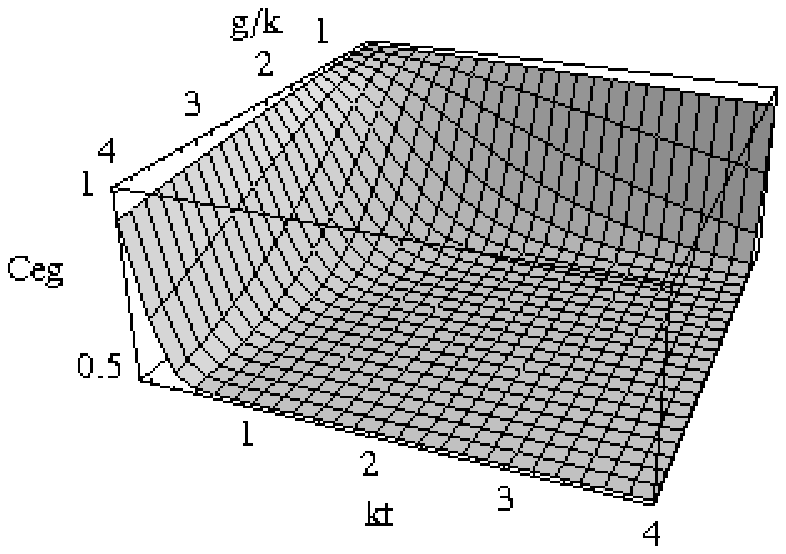} } b)
\resizebox{0.55\columnwidth}{!}{ \includegraphics{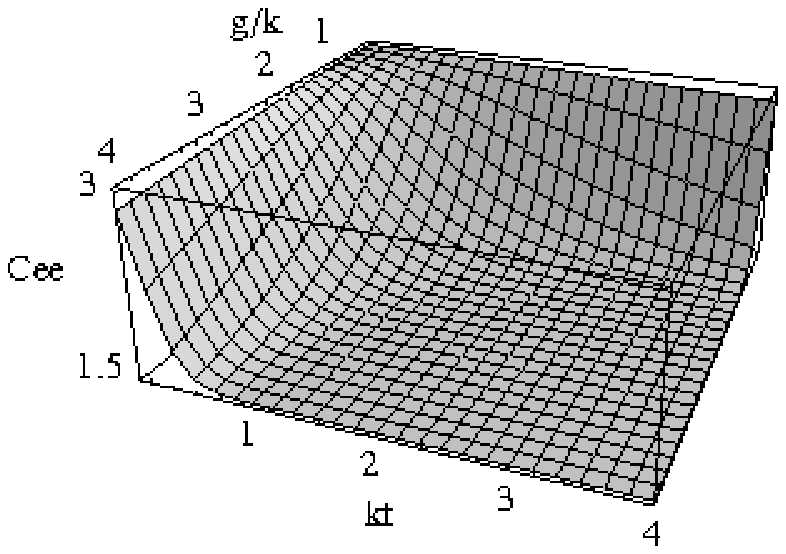} } c)
\caption{Correlation functions (a) $C_{gg}$, (b) $C_{eg}$, (c)
$C_{ee}$ vs. dimensionless time $kt$ and coupling constant $g/k$,
showing atomic bunching and anti-bunching.} \label{fig4}
\end{figure}

Its behavior is quite similar to the one of Fig.~\ref{fig2} for the
whole system purity. From~Eq.~(\ref{mx:gg_±}) we can derive the
single-atom density matrices and evaluate the probability to measure
one atom in the excited or ground state
\begin{equation}\label{Peg}
P_{{e,g}}(t)=\frac{1}{2}\left[1 \mp f_1(t) \right].
\end{equation}
Quite similar expressions hold for atoms prepared in any state of
the standard basis. We see that from measurements of the atomic
inversion $I(t)=p_g(t)-p_e(t)$ we can monitor the one-atom
decoherence function $f_1(t)$ as in \cite{SDOAL}. By rewriting the
atomic density matrix (\ref{mx:gg_±}) in the standard basis we
evaluate the joint probabilities
$P_{lm}(t)=\meanvalue{lm}{\hat{\rho}_{a}^{(e,g)}(t)}{lm}$ with
$\{l,m\}=\{e,g\}$. The corresponding correlation functions at a
given time $t$ are:
\begin{eqnarray}\label{corr-func}
C_{ee}(t) &=&\frac{3-4f_1(t)+f_2(t)}{2[1-f_1(t)]^2}\,\nonumber\\
C_{eg}(t) &=&C_{ge}(t)=\frac{1-f_2(t)}{2[1-f_1^2(t)]}\,\nonumber\\
C_{gg}(t)&=&\frac{3+4f_1(t)+f_2(t)}{2[1+f_1(t)]^2}
\end{eqnarray}
In Figs.~\ref{fig4}(a-c) we show the correlation functions
$C_{lm}(t)$ versus dimensionless time $kt$ and coupling constant
$\frac{g}{k}$.
At steady-state we see that $C_{gg}(t),C_{ee}(t)\mapsto 3/2$, that
is positive atom-atom correlation or bunching, whereas
$C_{eg}(t)\mapsto 1/2$, indicating negative correlation or
anti-bunching. These results generalize those in \cite{SDM}, which
can be derived from Eqs.(\ref{corr-func}) in the limit of negligible
dissipation $kt\ll 1$. We notice that from the joint atomic
probability $P_{eg}(t)=P_{ge}(t)=\frac{1}{8}[1-f_2(t)]$ one can
monitor the two-atom decoherence described by $f_2(t)$.

\section{Dynamics of entanglement and decoherence of the atomic subsystem}\label{sec:4}
\begin{figure}[h]
\resizebox{0.55\columnwidth}{!}{ \includegraphics{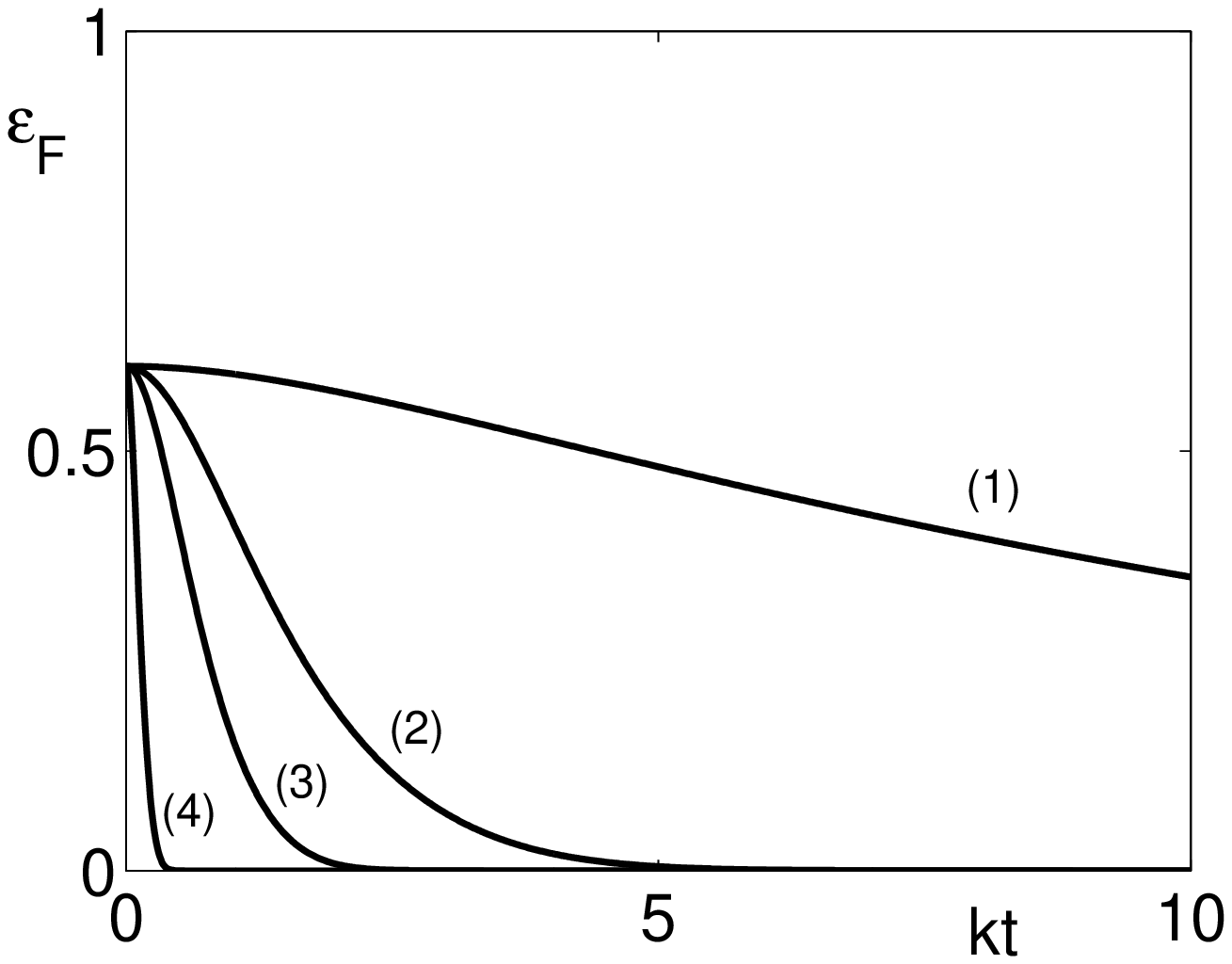} } a)
\resizebox{0.55\columnwidth}{!}{ \includegraphics{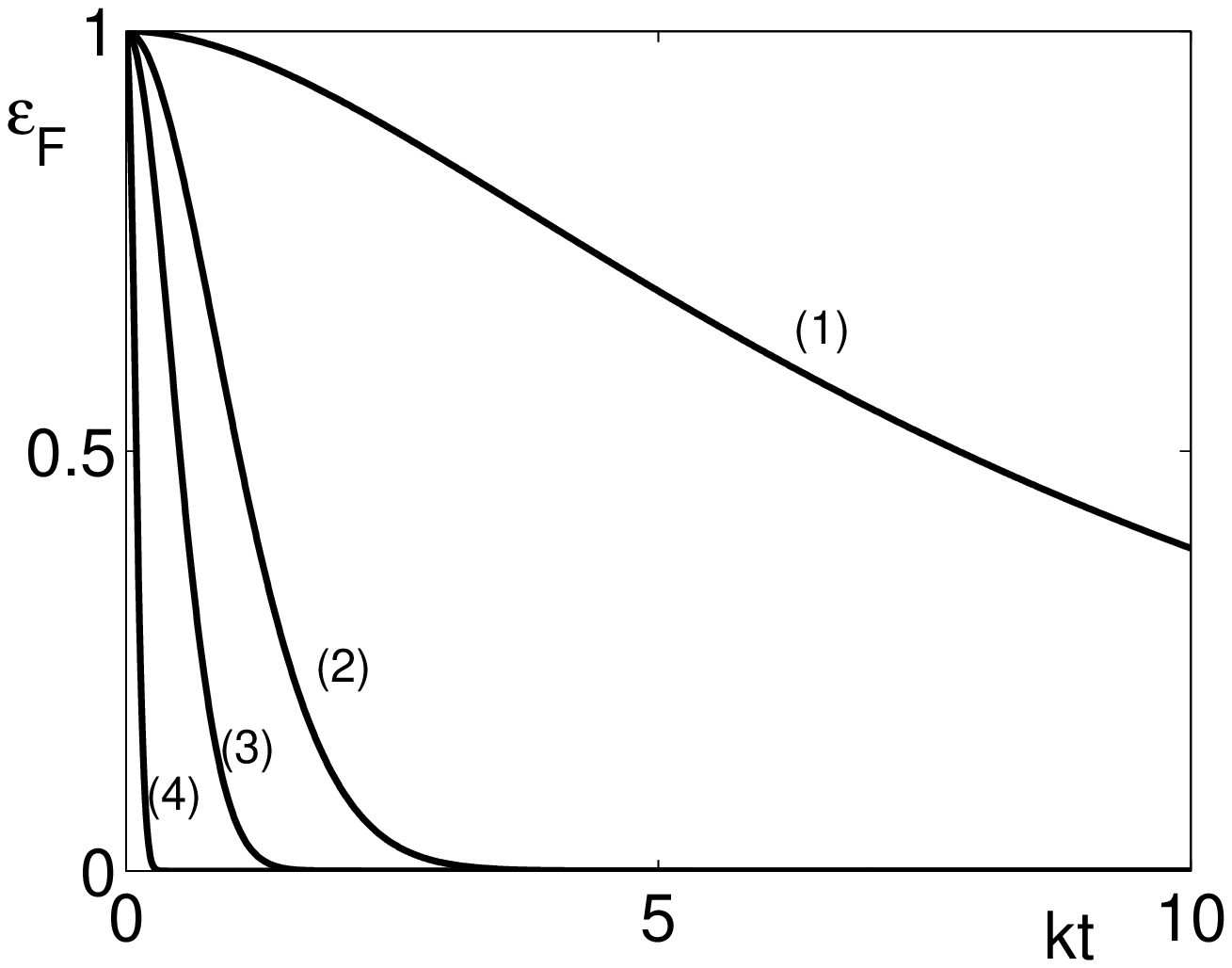} } b)
\caption{Entanglement of formation $\epsilon_F(t)$ as a function of
$kt$ and for different values of $g/k$: 0.1 (1), 0.5 (2), 1 (3), 5
(4). (a) atoms prepared in a partially entangled state
$\frac{|\Phi^+\rangle+e^{i\pi/4}|\Phi^-\rangle}{\sqrt{2}}$, (b)
atoms prepared in the Bell states $|\Phi^+\rangle$.} \label{fig5}
\end{figure}

\begin{figure}[h]
\resizebox{0.60\columnwidth}{!}{ \includegraphics{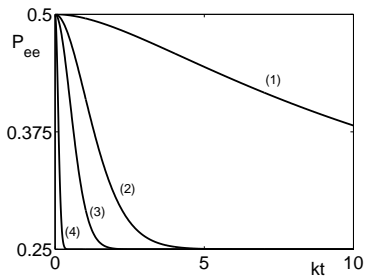} }
\caption{Atoms prepared in the Bell state $|\Phi^+\rangle$: joint
atomic probability $P_{ee}(t)$ vs. $kt$ for different values of
$g/k$: 0.1 (1), 0.5 (2), 1 (3), 5 (4).} \label{fig6}
\end{figure}

In Sec.~\ref{sec:3} we discussed the solution of the ME
(\ref{eq:ME3}) for both atoms prepared in the ground state as an
example of separable state. In order to describe also initially
entangled atoms now we consider the general solution derived in the
Appendix for atoms prepared in a superposition of Bell states
$|\psi_{a}(0)\rangle=\sum_{i=1}^4 c_i |\nu_i\rangle$, where the
coefficients $c_i$ are normalized as $\sum_{i=1}^4|c_i|^2=1$, and
$\{|\nu_i\rangle \}_{i=1,..,4}=\{|\Phi^+\rangle, |\Phi^-\rangle,
|\Psi^+\rangle, |\Psi^-\rangle\}$ is the Bell basis where
$|\Phi^\pm\rangle=\frac{|ee\rangle\pm |gg\rangle}{\sqrt{2}}$ and
$|\Psi^\pm\rangle=\frac{|eg\rangle\pm |ge\rangle}{\sqrt{2}}$.
Tracing over the atomic variables the solution for the whole density
operator (Eqs. (\ref{decomposition}) and (\ref{App:rhoijBell})) we
derive the cavity field density operator generalizing
Eq.~(\ref{rho_f}):
\begin{eqnarray}
\label{roF_BEll} &\hat{\rho}_f(t)&=\frac{1}{2}\bigg [|c_1+c_3|^2
\ketbra{-\alpha(t)}{-\alpha(t)}+\,\nonumber\\
&&+2(|c_2|^2+|c_4|^2)\ketbra{0}{0}+
|c_1-c_3|^2\ketbra{\alpha(t)}{\alpha(t)}\bigg] . \nonumber\\
\end{eqnarray}
For the mean photon number we obtain
$\langle\hat{N}\rangle(t)=(|c_1|^2+|c_3|^2)|\alpha(t)|^2$ that is
independent of the coefficients of states $|\Phi^-\rangle$ and
$|\Psi^-\rangle$. On the other hand, tracing over the field
variables we obtain the atomic density matrix. We report the
solution in the so called \textit{magic basis} \cite{Magic} that can
be obtained from the Bell basis simply multiplying $|\nu_2\rangle$
and $|\nu_3\rangle$ by the imaginary unit:
\begin{widetext}
\begin{eqnarray}
\label{roat_Bell} \rho_{a}^{mB}=
\left(\begin{array}{cccc} \frac{[1+f_2]|c_1|^2+[1-f_2]|c_3|^2}{2}
& -if_1c_1c_2^* & -i\frac{[1+f_2]c_1c_3^*+[1-f_2]c_1^*c_3}{2} & f_1c_1c_4^* \\
if_1c_1^*c_2 & |c_2|^2& f_1c_2c_3^* & ic_2c_4^*\\
 i\frac{[1+f_2]c_1^*c_3+[1-f_2]c_1c_3^*}{2} & f_1c_2^*c_3
 &  \frac{[1-f_2]|c_1|^2+[1+f_2]|c_3|^2}{2}& if_1c_3c_4^* \\
f_1c_1^*c_4 & -i c_2^*c_4 &  -i f_1c_3^*c_4 & |c_4|^2
\end{array}\right) , \,\nonumber\\
\end{eqnarray}
\end{widetext}
where we omitted the time dependence for brevity. To evaluate the
entanglement properties of the atomic subsystem we consider the
entanglement of formation $\epsilon_F(t)$~\cite{EOF} defined as
\begin{eqnarray}
\label{EFdef}
\epsilon_{F}(t)&=&-\frac{1-\sqrt{1-C^2(t)}}{2}\log_2\frac{1-\sqrt{1-
C^2(t)}}{2}+\,\nonumber\\
&-&\frac{1+\sqrt{1-C^2(t)}}{2}\log_2\frac{1+\sqrt{1-C^2(t)}}{2} ,
\end{eqnarray}
where $C(t)$ is the concurrence that can be evaluated as
$C(t)=max\{0,\Lambda_4(t)-\Lambda_3(t)-\Lambda_2(t)-\Lambda_1(t)\}$
($\Lambda_i$ are the square roots of the eigenvalues of the non
hermitian matrix $\rho_{a}^{mB}(t) (\rho_{a}^{mB}(t))^*$ taken in
decreasing order). From Eq.~(\ref{roat_Bell}) we can derive the
probabilities for joint atomic measurements in the standard basis
\begin{eqnarray}\label{jointprob}
P_{ee,gg}(t)&=&\frac{1}{4}\big[|c_1|^2(1+f_2(t))+2|c_2|^2\,\nonumber\\
&+& |c_3|^2(1-f_2(t)) \pm 4f_1(t)Re(c_1c_2^*) \big]\,\nonumber\\
P_{eg,ge}(t)&=&-\frac{1}{4}\big[ |c_1|^2(1-f_2(t)) + 2|c_4|^2\,\nonumber\\
&+&|c_3|^2(1+f_2(t)) \pm 4f_1(t)Re(c_3c_4^*) \big].
\end{eqnarray}
Also, we can derive the density matrix corresponding to a single
atom and obtain the atomic probabilities generalizing
Eq.~(\ref{Peg}):
\begin{eqnarray}
\label{peg_Bell} P_{e,g}(t)=\frac{1}{2}\left[ 1\pm 2 f_1(t)
Re(c_1^*c_2+c_3^*c_4) \right].
\end{eqnarray}
\indent First we consider the case of a superposition of Bell states
$|\Phi^\pm\rangle$ (i.e., $c_1=a, c_2=be^{i\theta},c_3=c_4=0$ with
$a,b$ real numbers). The initial atomic state $|gg\rangle$
($|ee\rangle$) can be obtained if $a=b=\frac{1}{\sqrt{2}}$ and
$\theta=\pi$ ( $\theta=0$). We find that the concurrence $C(t)$
vanishes for any time and every value of $g/k$, so that it is not
possible to entangle the atoms. In the case of atoms prepared in a
partially entangled state we find that the entanglement of formation
can only decrease during the system evolution as shown for example
in Fig.~\ref{fig5}(a) in the case $a=b=\frac{1}{\sqrt{2}}$ and
$\theta=\pi/4$. We also see that the progressive loss of
entanglement is faster for large values of the parameter $g/k$. For
atoms prepared in the maximally entangled state $|\Phi^+\rangle$
(i.e., $\theta=0, a=1, b=0$) we derive that the concurrence simply
reduces to $f_2(t)$ that also describes the whole system
decoherence. This important point will be discussed later. In
Fig.~\ref{fig5}(b) we show the entanglement of formation as a
function of dimensionless time $kt$. For atoms prepared in the
maximally entangled state $|\Phi^-\rangle$ (i.e.,
$\theta=0,a=0,b=1$) the concurrence is always maximum ($C(t)=1$). In
fact, we can see from Eq.(\ref{roat_Bell}) that the atomic density
matrix is always the one of the initial state, as expected because
$|\Phi^-\rangle$ is a linear combination of the invariant states
$|+-\rangle$
and $|-+\rangle$.\\
\indent Starting from a superposition of Bell states
$|\Psi^\pm\rangle$ (i.e., $c_3=a, c_4=b e^{i\theta}, c_1=c_2=0$) we
obtain analogous results. In particular entanglement cannot be
generated for atoms prepared in states $|eg\rangle$ and
$|ge\rangle$, for the state $|\Psi^+\rangle$ the concurrence is
given by $f_2(t)$, and the entanglement of state $|\Psi^-\rangle$ is
preserved during system evolution. We find analogous results also
for the concurrence of atoms prepared in a superposition of
$|\Phi^-\rangle$ and $|\Psi^+\rangle$ (i.e., $c_2=a, c_3=b
e^{i\theta}, c_1=c_4=0$) or in a superposition of $|\Phi^+\rangle$
and $|\Psi^-\rangle$
(i.e., $c_1=a, c_4=b e^{i\theta}, c_2=c_3=0$).\\
\indent Let us summarize and discuss the main results in the case of
atoms prepared in entangled states. If we consider a superposition
of states $|\Phi^-\rangle$ and $|\Psi^-\rangle$ (i.e., $c_2=a, c_4=b
e^{i\theta}, c_1=c_3=0$) we find that the atomic density matrix of
Eq.(\ref{roat_Bell}) does not evolve. Actually, the atomic subspace
spanned by $|\Phi^-\rangle$ and $|\Psi^-\rangle$ coincides with the
time-invariant subspace spanned by $|+-\rangle$ and $|-+\rangle$, so
that it remains protected from dissipation during the system
evolution. It provides an example of Decoherence Free Subspace
(DFS)\cite{DFS}. Atomic entanglement injected in the system can thus
remain available for long storage times for applications in quantum
information
processing \cite{Minns}.\\
\indent If we consider a superposition of states $|\Phi^+\rangle$
and $|\Psi^+\rangle$ (i.e., $c_1=a, c_3=be^{i\theta}, c_2=c_4=0$)
the concurrence is given by:
\begin{equation}
\label{Conc2} C(t)=\sqrt{a^4+b^4-2a^2b^2 \cos(2\theta)} f_2(t)
\end{equation}
and we find the remarkable result that the initial entanglement is
progressively reduced by the decoherence function $f_2(t)$. Note
that $C(t)=f_2(t)$ for states $|\Phi^+\rangle$, $|\Psi^+\rangle$,
and any superposition as $a|\Phi^+\rangle\pm ib|\Psi^+\rangle$. To
understand this point let us consider the specific example of the
initial state $|0\rangle\otimes|\Phi^+\rangle$. The evolved density
operator of the whole system is
\begin{eqnarray}\label{evolvedrho}
\hat{\rho}(t)=\frac{1}{2}\big\{
\ketbra{-\alpha(t)}{-\alpha(t)}\otimes\ketbra{++}{++}+\,\nonumber\\
\ketbra{\alpha(t)}{\alpha(t)} \otimes\ketbra{--}{--}+\,\nonumber\\
\frac{f_{2}(t)}{e^{-2|\alpha(t)|^2}} \big[
\ketbra{-\alpha(t)}{\alpha(t)}\otimes\,\nonumber\\\ketbra{++}{--}
+ \ketbra{\alpha(t)}{-\alpha(t)}\otimes\ketbra{--}{++}\big] \big\}.\,\nonumber\\
\end{eqnarray}
In the limit, $kt\ll1$, of short time and/or negligible dissipation,
where $f_2(t)\simeq e^{-2|\tilde{\alpha}(t)|^2}$ with
$\tilde{\alpha}(t)=igt$, the system evolves into a pure cat-like
state where the atoms are correlated with coherent states
\begin{equation}\label{evolvedket}
|0\rangle\otimes\frac{|++\rangle + |--\rangle}{\sqrt{2}} \mapsto
\frac{|-\tilde{\alpha}\rangle\otimes|++\rangle +
|\tilde{\alpha}\rangle\otimes|--\rangle}{\sqrt{2}}.
\end{equation}
For longer times/larger dissipation, the system coherence decays as
described by the function $f_2(t)$. Let us now consider the atomic
dynamics disregarding the field subsystem. The reduced atomic
density operator is
\begin{eqnarray}\label{evolvedrhoatoms}
\hat{\rho}_a(t)=\frac{1}{2}\big[
|++\rangle\langle++|+|--\rangle\langle--|\,\nonumber\\
+ f_{2}(t)\big(|++\rangle\langle--|+|--\rangle\langle++| \big)\big].
\end{eqnarray}
Hence, as the quantum coherence reduces, simultaneously the atoms
lose their inseparability, and the state becomes maximally mixed (in
the relevant subspace). In fact, the atomic purity is given by
$\mu_a(t)=\frac{1+f_2^2(t)}{2}$. Hence the time evolution of
decoherence, concurrence, and purity is described by the function
$f_2(t)$, which can be monitored via a measurement of joint atomic
probabilities (see Eq. (\ref{jointprob}))
\begin{eqnarray}\label{Pee}
P_{ee}(t)&=&P_{gg}(t)=\frac{1}{4}\left[1 + f_2(t)
\right]\,\nonumber\\
P_{eg}(t)&=&P_{ge}(t)=\frac{1}{4}\left[1 - f_2(t) \right].
\end{eqnarray}
The probability $P_{ee}(t)$ is shown in Fig.~\ref{fig6} as a
function of $kt$ for different values of the ratio $g/k$. For
$kt\ll1$, when the whole system is in the cat-like state
(\ref{evolvedket}), $f_2(t)$ quadratically decreases as
$\exp{(-2g^2t^2)}$, independent of the dissipative rate $k$. The
subsequent behavior is approximately an exponential decay whose
start and rate depend on the atom-cavity field coupling. For
$g/k\gtrsim 0.5$, that is also below the strong coupling regime, we
can introduce a decoherence and disentanglement rate
\begin{equation}
\gamma_D \simeq k |\alpha^{SS}| = 2g,
\end{equation}
that is again independent of (and faster than) $k$. A physical
interpretation of this result is that the more coupled the two atoms
are to the dissipative cavity mode, the more effective becomes the
decay of both the environment-induced decoherence and the initial
entanglement. For $g/k\ll 1$, that is in a weak coupling regime, the
exponential decay of coherence and concurrence starts later and its
rate, $\gamma_D \simeq 8g^2/k$, is slower than the dissipative rate $k$.\\
\indent We have shown that under full resonance conditions it is not
possible to generate or increase the initial atomic entanglement.
This can be explained looking at the initial requirements that allow
us to write Eq.(\ref{App:16operators}). In particular, the strong
driving condition $\Omega \gg g$, the resonance condition $\delta=0$
and the choice of negligible atomic decays $\gamma=0$ are necessary
to obtain an independent set of equations for the operators
$\hat{\rho}_{ij}$ and to exactly solve the system dynamics. In
section \ref{sec:6} we will show that removing the strong driving
condition it is possible to slightly entangle the atoms. On the
other hand, it can be shown \cite{CL} that with off-resonant
atoms-cavity field interaction it is possible to generate maximally
entangled atomic states also under strong driving conditions. Also
we recall that, in the resonant case and without driving field, two
atoms prepared in a separable state can partially entangle by
coupling to a thermal cavity field~\cite{Kim}.\\

\section{Conditional generation of states}\label{sec:5}
In this section we seek information about the states of one of the
two subsystems conditioned by a projective measurement on the other one.\\
\indent If the system is implemented in the optical domain and the
cavity field is accessible to measurements, a null measurement by a
on/off detector implies the generation of a maximally entangled
atomic Bell state~\cite{PavelOptical}. In the case of atoms prepared
in state $\ket{gg}$ (or $\ket{ee}$), the atomic conditioned state
will be the Bell state $\ket{\Phi^-}$ (see (\ref{gg0})).
Analogously, for initial atomic state $\ket{eg}$ or $\ket{ge}$ the
atomic conditioned state will be $\ket{\Psi^-}$.\\
\indent Now we consider the evolution of the field subsystem
conditioned by a projective atomic measurement on the bare basis
$\{\ket{ee},\ket{eg},\ket{ge},\ket{gg}\}$. Starting e.g. from the
initial state $|0\rangle\otimes|gg\rangle$, the cavity field will be
in the conditioned states at a given time $t$ (omitted for brevity
in this section):
\begin{eqnarray}\label{eq:rho_condiz_eegg}
\hat{\rho}_{f,(ee,gg)}&=&\frac{1}{2(3+f_2\mp 4f_1)}\bigg [ \ketbra{-\alpha}{-\alpha}+\ketbra{\alpha}{\alpha}+\,\nonumber\\
&+&4\ketbra{0}{0}+f_2e^{2|\alpha|^2}\bigg ( \ketbra{\alpha}{-\alpha}+\ketbra{-\alpha}{\alpha}\bigg )\,\nonumber\\
&\mp& 2f_1e^{\frac{|\alpha|^2}{2}}\bigg (
\ketbra{0}{-\alpha}+\ketbra{0}{\alpha}+\,\nonumber\\
&+&\ketbra{-\alpha}{0}+\ketbra{\alpha}{0}\bigg
) \bigg ]\,\nonumber\\
\hat{\rho}_{f,(eg,ge)}&=&\frac{1}{2(1-f_2)}\bigg [
\ketbra{-\alpha}{-\alpha}+\ketbra{\alpha}{\alpha}\,\nonumber\\
&-&f_2e^{2|\alpha|^2}\bigg (
\ketbra{\alpha}{-\alpha}+\ketbra{-\alpha}{\alpha}\bigg ) \bigg ].
\end{eqnarray}
Note that in the limit $kt\ll 1$ the conditioned field state is a
Schr\"odinger-cat-like state
\begin{eqnarray}
\label{eq:psi_gatto}
\ket{\psi}_{f,(ee,gg)}&=&\frac{ \ket{-\tilde{\alpha}}\mp 2\ket{0}+
\ket{\tilde{\alpha}}}{\sqrt{2(e^{-2|\tilde{\alpha}|^2}\mp4e^{-|\tilde{\alpha}|^2/2}+3)}}\,\nonumber\\
\ket{\psi}_{f,(eg,ge)}&=&\frac{
\ket{-\tilde{\alpha}}-\ket{\tilde{\alpha}}}{\sqrt{2(1-e^{-2|\tilde{\alpha}|^2})}}
\end{eqnarray}
where $\tilde{\alpha}(t)=igt$.\\
The Wigner functions representing the states
(\ref{eq:rho_condiz_eegg}) in phase space at a given time, are:
\begin{eqnarray}\label{eq:wigners}
W_{f,(ee,gg)}(\beta)&=&\frac{2e^{-2|\beta|^2}}{\pi(3+f_2\mp 4f_1)}\bigg [2 + \,\nonumber\\
&+&e^{-2|\alpha|^2}\cosh(4|\alpha|Im\beta)\,\nonumber\\
&+&f_2e^{2|\alpha|^2}\cos(4|\alpha|Re\beta)\,\nonumber\\
&\mp&4f_1\cosh(2|\alpha|Im\beta)\cos(2|\alpha|Re\beta)\bigg ]\,\nonumber\\
W_{f,(eg,ge)}(\beta)&=&\frac{2e^{-2|\beta|^2}}{\pi(1-f_2)}\bigg [e^{-2|\alpha|^2}\cosh(4|\alpha|Im\beta)\,\nonumber\\
&-&f_2e^{2|\alpha|^2}\cos(4|\alpha|Re\beta)\bigg ].
\end{eqnarray}
\begin{figure}[h]
\resizebox{0.90\columnwidth}{!}{ \includegraphics{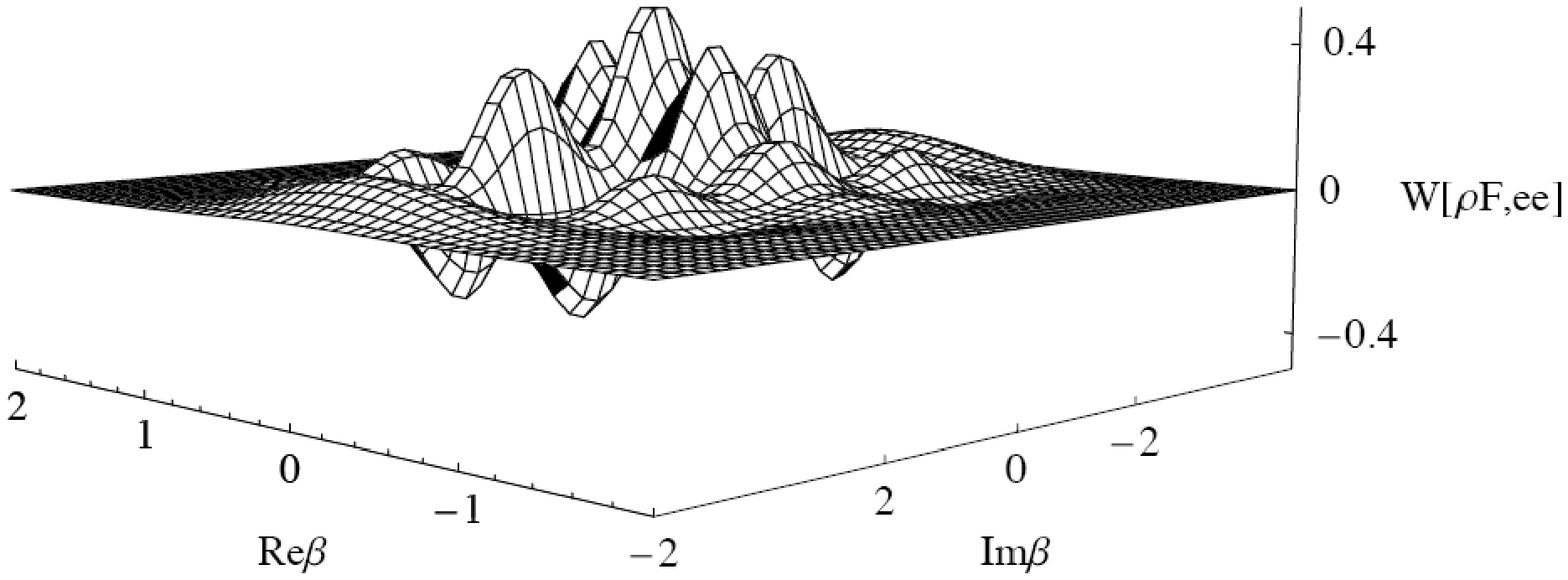} } a)\\
\resizebox{0.90\columnwidth}{!}{ \includegraphics{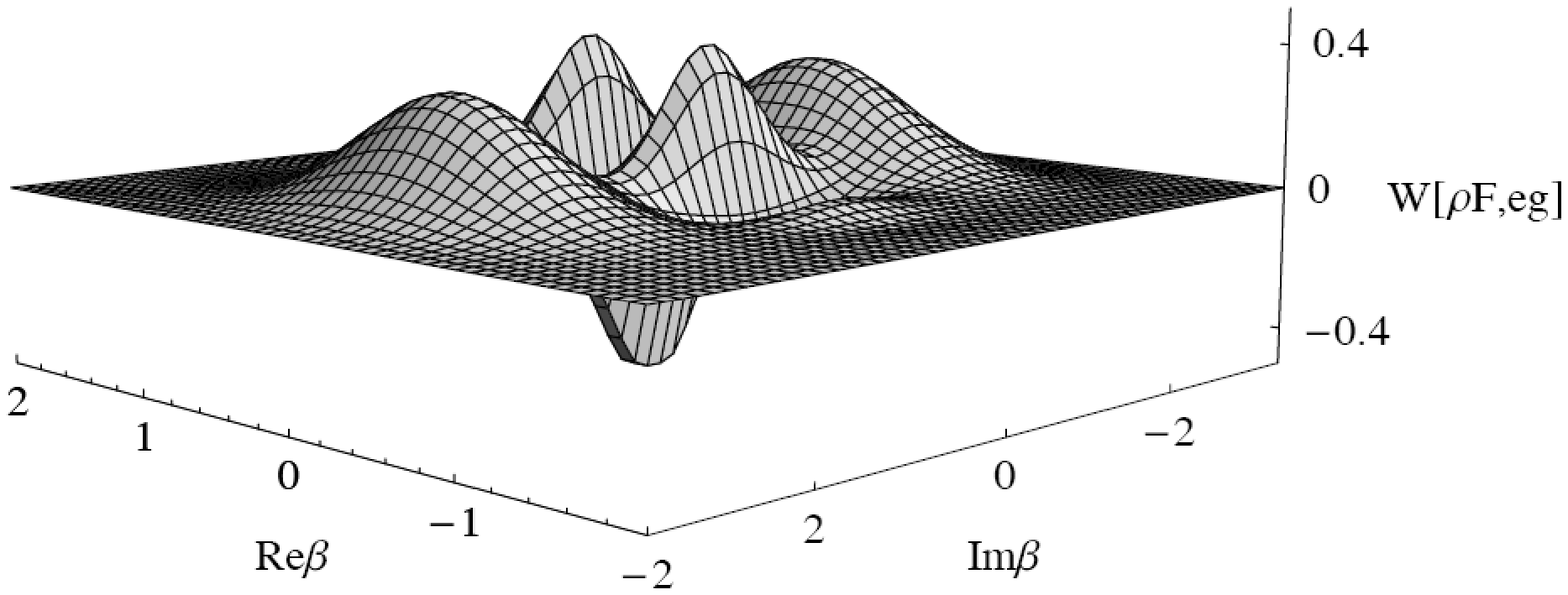} } b)\\
\caption{Wigner function of the cavity field
$W[\hat{\rho}_{f,(ee)}]$ (a) and $W[\hat{\rho}_{f,(eg)}]$ (b), for
$kt=0.05$, and for $\frac{g}{k}=80$ (a) and $\frac{g}{k}=40$ (b). }
\label{fig7}
\end{figure}
In Fig.~\ref{fig7} we illustrate two of the Wigner functions
(\ref{eq:wigners}) in the transient $kt\ll 1$ and in the strong
coupling regime $g\gg k$. At steady state the Wigner functions are
positive, corresponding to the states:
\begin{eqnarray}\label{eq:campo_cond_SS}
\hat{\rho}^{SS}_{F,(ee,gg)}&=&\frac{1}{6}\left [ \ketbra{-\alpha^{SS}}{-\alpha^{SS}}
+\ketbra{\alpha^{SS}}{\alpha^{SS}}+4\ketbra{0}{0}\right ]\,\nonumber\\
\hat{\rho}^{SS}_{F,(eg,ge)}&=&\frac{1}{2}\left [
\ketbra{-\alpha^{SS}}{-\alpha^{SS}}+\ketbra{\alpha^{SS}}{\alpha^{SS}}\right
]\end{eqnarray} where $\alpha^{SS}=i\frac{2g}{k}$.

\section{Numerical results}\label{sec:6}
To confirm our theoretical analysis as well as to investigate system
dynamics without the strong driving condition and including the
effect of atomic decay, where analytical results are not available,
we numerically solve, by Monte Carlo Wave Function (MCWF) method
\cite{MCWF}, the ME (\ref{eq:MEIP}) in the resonant case $\delta=0$,
that we rewrite in the Lindblad form:
\begin{eqnarray}
\label{eq:maseqLFNUM}\dot{\hat{\rho}}_I& = & -
\frac{i}{\hbar}(\hat{\mathcal{H}}_{e}
\hat{\rho}_I-\hat{\rho}_I\hat{\mathcal{H}}_{e}^{\dagger})
+\sum_{i=1}^3 \hat{C}_i\hat{\rho}_I\hat{C}_i^{\dagger}
\end{eqnarray}
where the non-Hermitian effective Hamiltonian
$\hat{\mathcal{H}}_{e}$ is given by
\begin{equation}
\label{NorHermitianHNUM}\hat{\mathcal{H}}_{e}  =
\frac{\hat{\mathcal{H}}_{I}}{g} -
\frac{i\hbar}{2}\sum_{i=1}^3\hat{C}_i^{\dagger}\hat{C}_i,
\end{equation}
the Hamiltonian $\hat{\mathcal{H}}_{I}$ is that of
Eq.(\ref{eq:IP_H}) for $\delta=0$ , and the collapse operators are
$\hat{C}_{1,2}=\sqrt{\tilde{\gamma}}\hat{\sigma}_{1,2}$,
$\hat{C}_3=\sqrt{\tilde{k}}\hat{a}$. We have introduced the scaled
time $\tilde{t}=gt$ so that the relevant dimensionless system
parameters are:
\begin{equation}
\label{Dimpar} \tilde{\Omega}=
\frac{\Omega}{g}\hspace{0.5cm}\tilde{k}=
\frac{k}{g}\hspace{0.5cm}\tilde{\gamma}=\frac{\gamma}{g}.
\end{equation}
The system dynamics can be simulated by a suitable number $N_{tr}$
of trajectories, i.e. stochastic evolutions of the whole system wave
function $|\psi_j(\tilde{t})\rangle$ ($j=1,2,...N_{tr}$). Therefore,
the statistical operator of the whole system can be approximated by
averaging over the $N_{tr}$ trajectories, i.e.,
$\hat{\rho}_I(\tilde{t})\cong
\frac{1}{N_{tr}}\sum_{i=j}^{N_{tr}}|\psi_j(\tilde{t})\rangle\langle\psi_j(\tilde{t})|$.\\
\indent First we consider negligible atomic decay $\tilde{\gamma}=0$
to confirm the analytical solutions and to evaluate the effect of
the driving parameter $\tilde{\Omega}$. For numerical convenience we
consider the case $\tilde{k}=1$ so that the steady state mean photon
number assumes small enough values. We consider the atoms prepared
in the maximally entangled states $|\Phi^{\pm}\rangle$. We recall
that for the state $|\Phi^{+}\rangle$ the theoretical mean photon
number is $\langle\hat{N}(\tilde{t})\rangle=|\alpha(\tilde{t})|^2$,
the atomic populations $P_{e,g}(\tilde{t})=0.5$, the atomic purity
$\mu_a(\tilde{t})=\frac{1+f_2^2(\tilde{t})}{2}$, and the
entanglement of formation $\epsilon_F(t)$  is given by (\ref{EFdef})
where the concurrence $C(t)$ coincides with $f_2(t)$.
\begin{figure}[h]
a)\resizebox{0.45\columnwidth}{!}{ \includegraphics{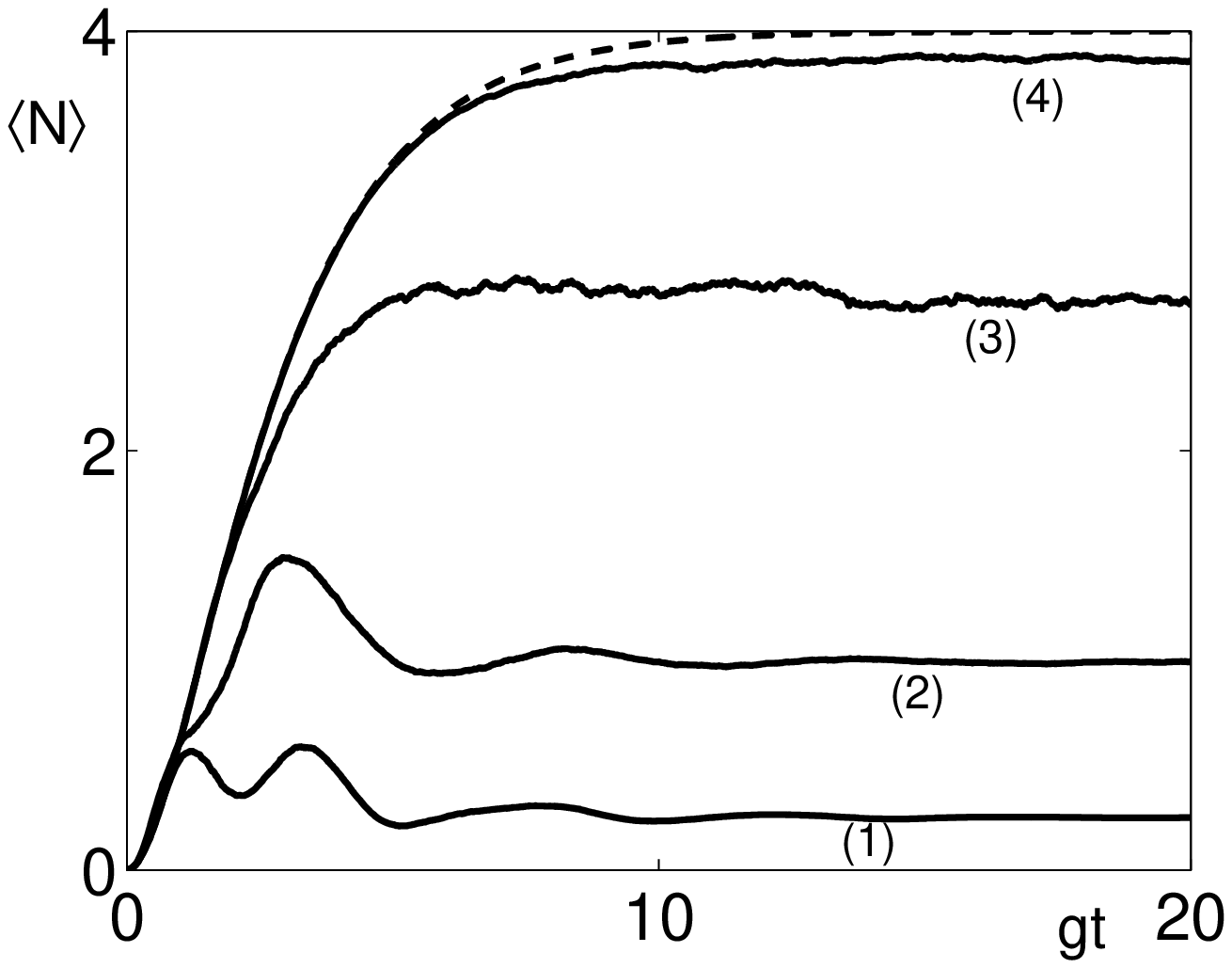}}
b)\resizebox{0.45\columnwidth}{!}{ \includegraphics{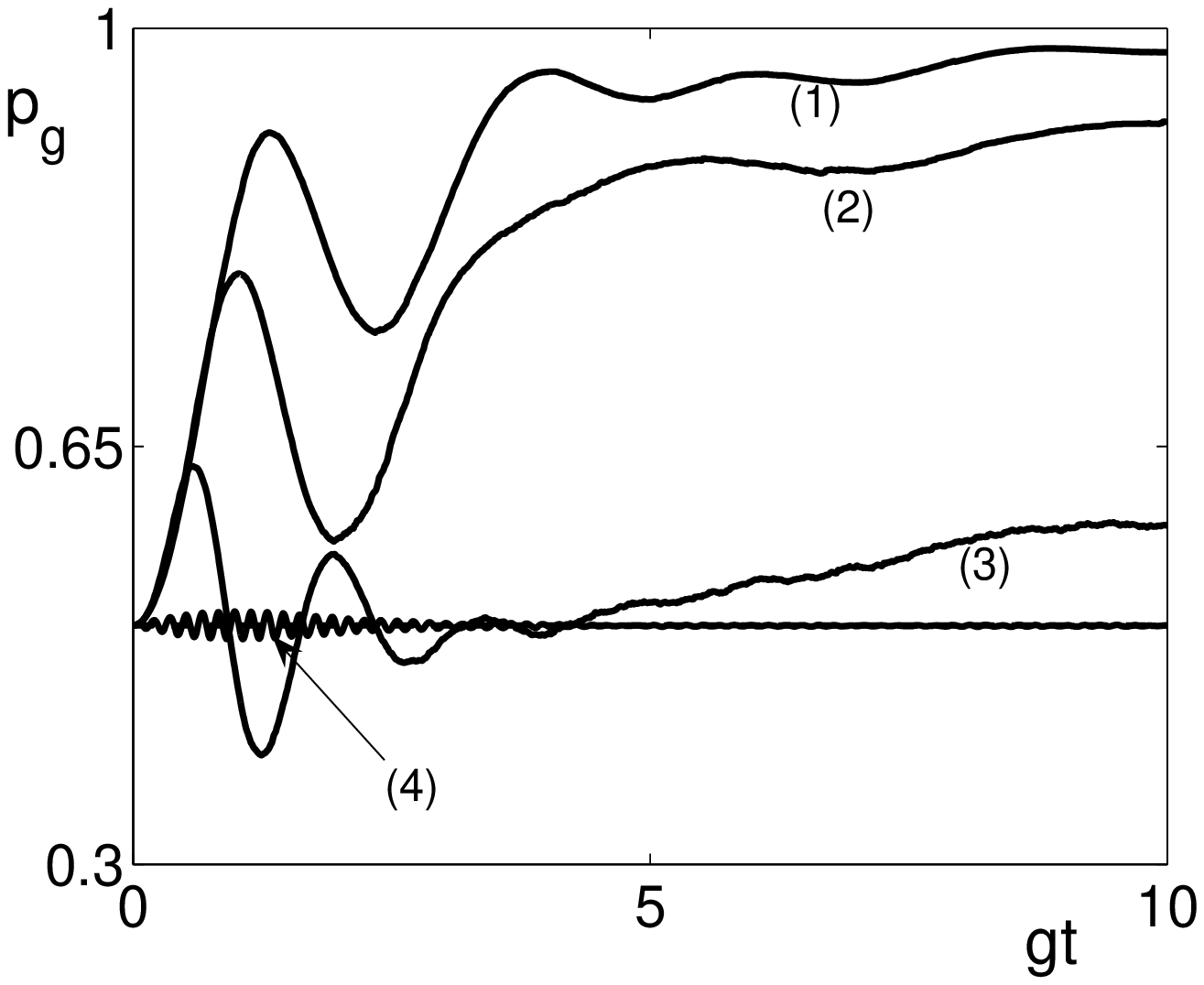}}\\
c)\resizebox{0.45\columnwidth}{!}{ \includegraphics{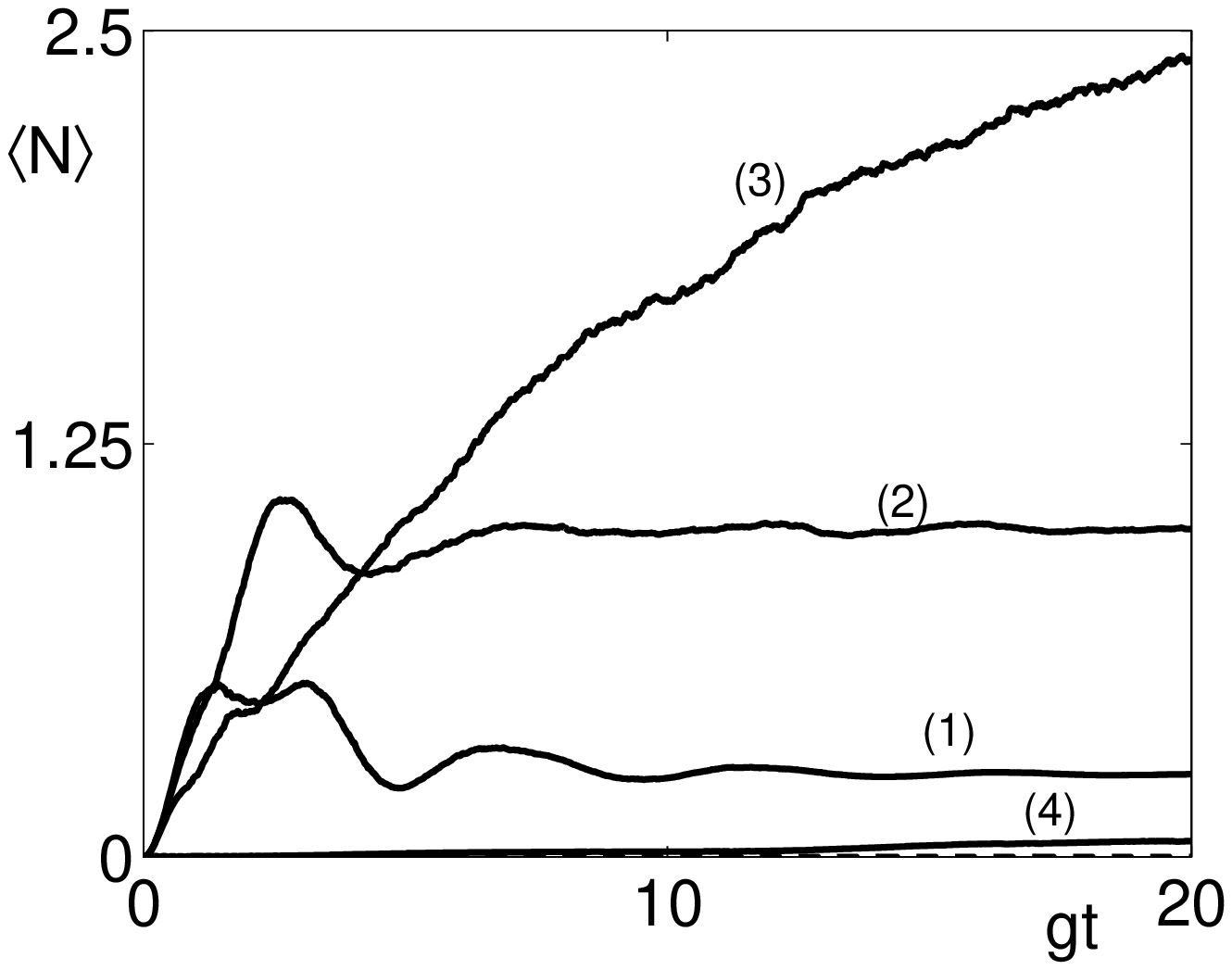}}
d)\resizebox{0.45\columnwidth}{!}{ \includegraphics{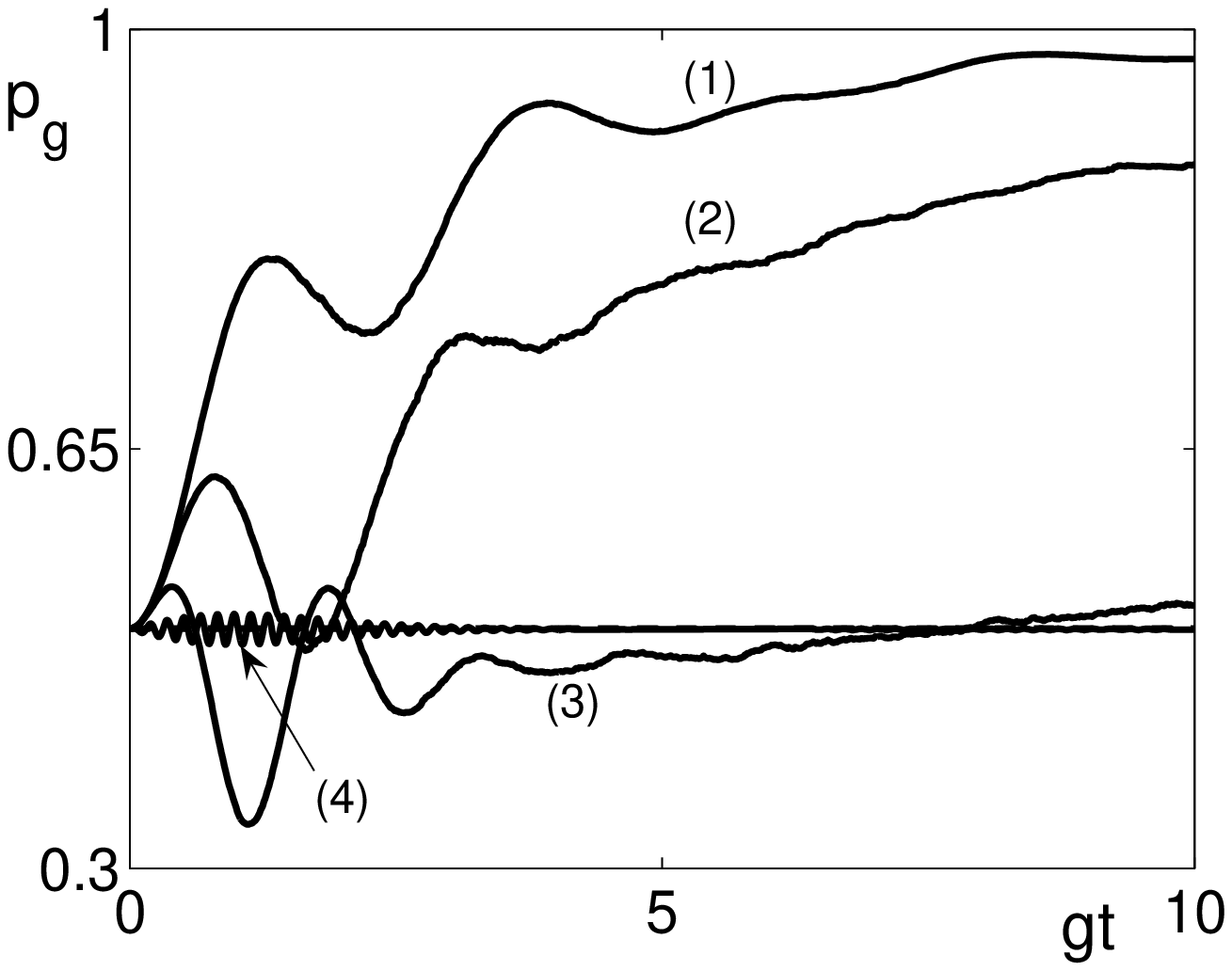}}\\
\caption{ Effect of driving parameter $\tilde{\Omega}$ for
negligible atomic decay $\tilde{\gamma}=0$, $\tilde{k}=1$, atoms
prepared in the Bell state $|\Phi^+\rangle$ ((a),(b)) and
$|\Phi^-\rangle$ ((c),(d)), for values of $\tilde{\Omega}$: $0.5$
(1), $1$ (2), $2$ (3), $20$ (4). The theoretical functions are the
dashed lines.  We show in (a),(c) the mean photon number
$\langle\hat{N}(\tilde{t})\rangle$, and in (b),(d) the atomic
probability $p_{g}(\tilde{t})$ . The number of trajectories is
$N_{tr}=500$.}\label{fig:8}
\end{figure}
In Fig.~\ref{fig:8} we show e.g. the mean photon number and the
atomic probability $p_g(t)$. In the strong driving limit,
$\tilde{\Omega}=20$, we find an excellent agreement with the
predicted theoretical behavior. We note that we simulated the system
dynamics without the RWA approximation so that $p_{g}(\tilde{t})$
exhibits oscillations due to the driving field. We remark that the
entanglement of formation in the case of state $|\Phi^+\rangle$
evolves almost independently of parameter $\tilde{\Omega}$ and
becomes negligible after times $gt \approx 2$. In the case of state
$|\Phi^-\rangle$ $\epsilon_F(t)$ decays in a similar way for small
values of $\tilde{\Omega}$, but it remains close to 1 for large
enough values of the driving parameter as predicted in our analysis.\\
\indent Finally, we consider the effect of the atomic decay. For
example, we consider the atoms prepared in the Bell state
$|\Phi^-\rangle$ in the strong driving limit and for the cavity
field decay rate $\tilde{k}=1$. In Fig.~\ref{fig:9} we show the mean
photon number and the entanglement of formation. We see that for
$\tilde{\gamma}$ up to $10^{-3}$ the effect of atomic decays is
negligible and the results of our treatment still apply. For larger
decay rates the atomic dynamics becomes no more restricted within
the decoherence free subspace.
\begin{figure}[h]
a)\resizebox{0.45\columnwidth}{!}{ \includegraphics{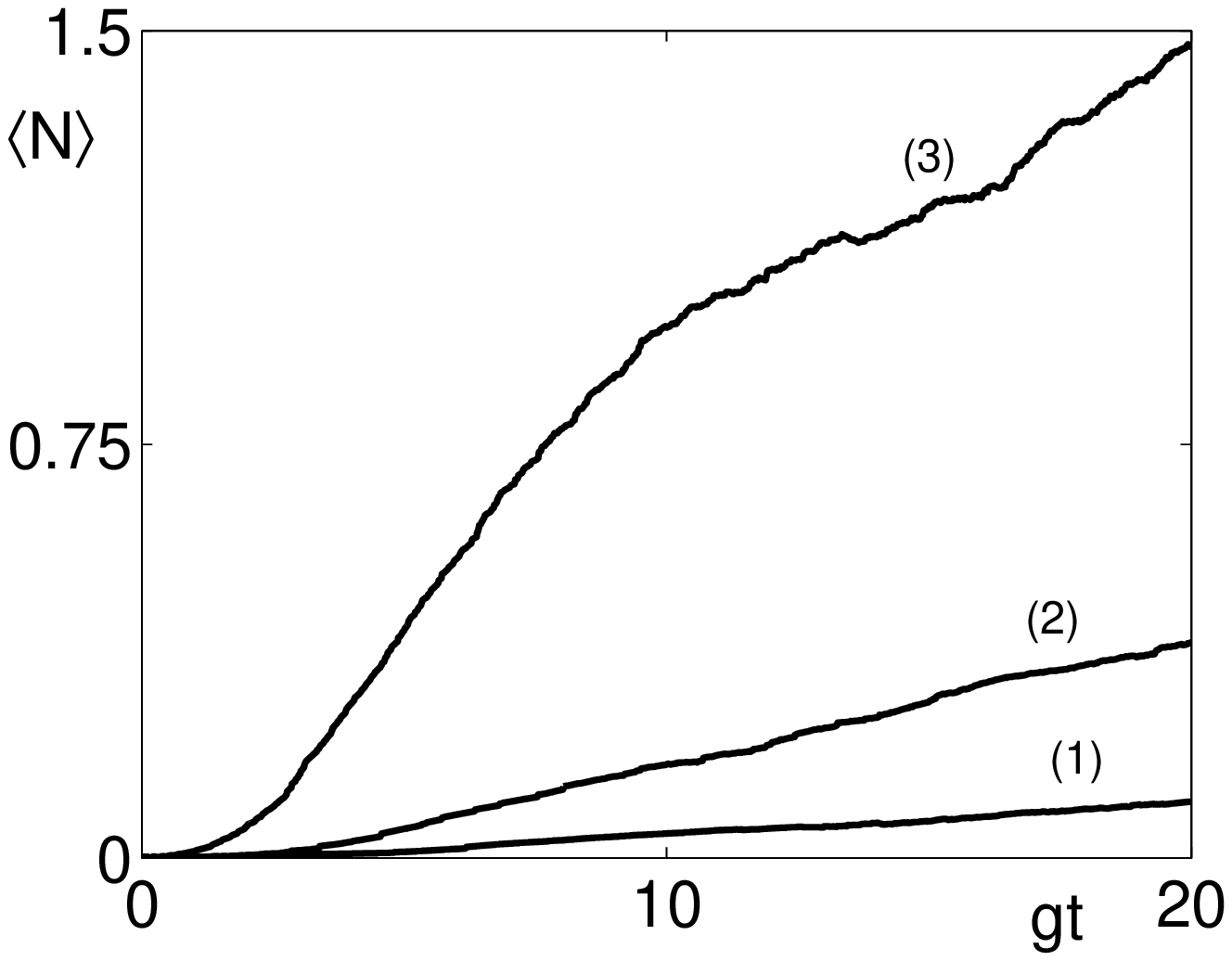}}
b)\resizebox{0.45\columnwidth}{!}{ \includegraphics{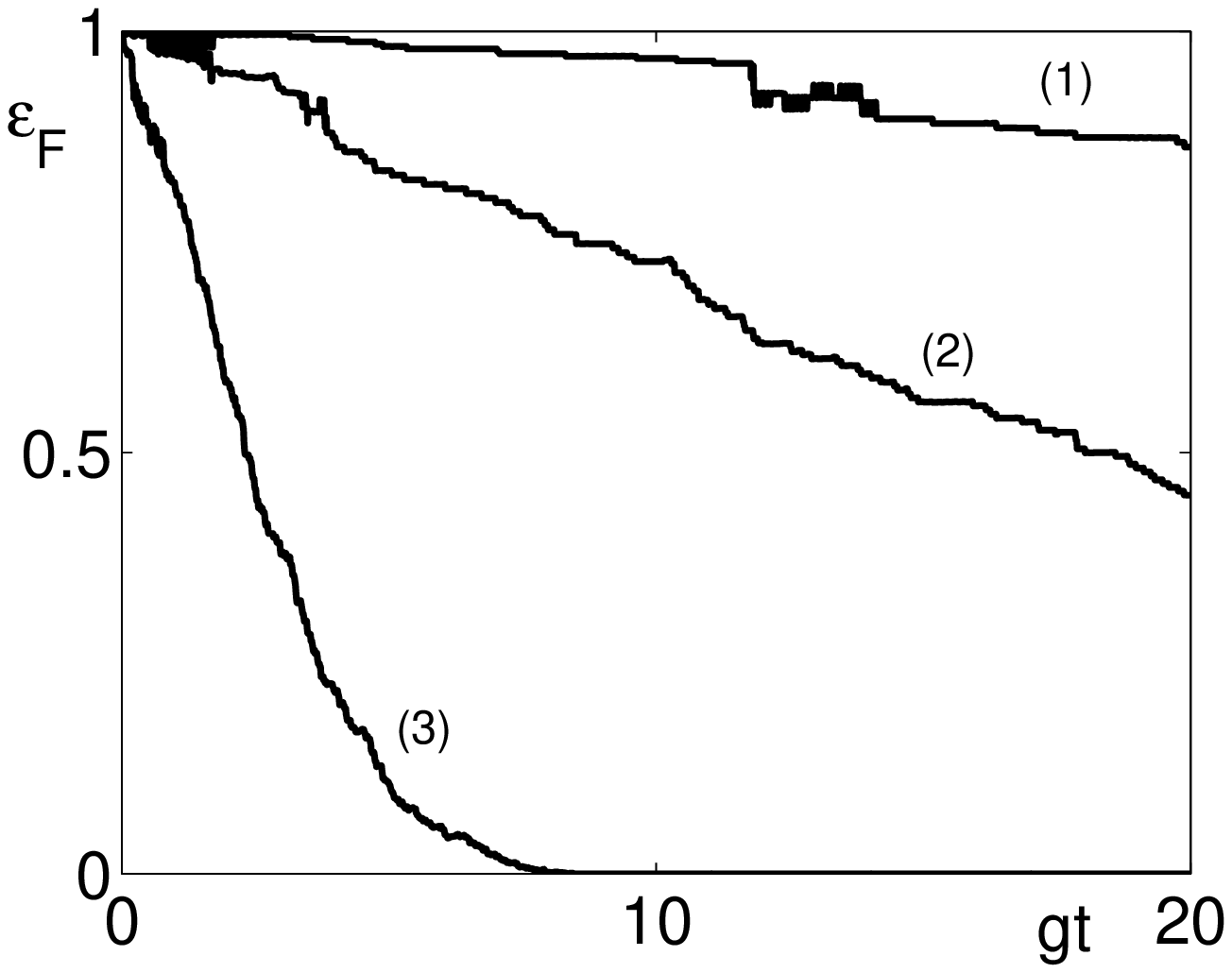}}\\
\caption{ Effect of atomic decay $\tilde{\gamma}$ in the strong
driving condition  $\tilde{\Omega}=20$, atoms prepared in the Bell
state $|\Phi^-\rangle$ and $\tilde{k}=1$: $\tilde{\gamma}=0.001$(1),
$0.01$ (2), $0.1$ (3). (a) Mean photon number
$\langle\hat{N}(\tilde{t})\rangle$, (b) Entanglement of formation
$\epsilon_F(\tilde{t})$. The number of trajectories is
$N_{tr}=500$.}\label{fig:9}
\end{figure}

\section{Conclusions}\label{sec:7}
We have described the dynamics of a system where a pair of two-level
atoms, strongly driven by an external coherent field, are resonantly
coupled to a cavity field mode that is in contact with an
environment. There are different available or forthcoming routes to
the implementation of our model. In the microwave regime of cavity
QED pairs of atoms excited to Rydberg levels cross a high-Q
superconductive cavity with negligible spontaneous emission during
the interaction \cite{Haroche-Raimond}. Difficulties may arise due
to the (ideal) requirements on atomic simultaneous injection, equal
velocity, and equal coupling rate to the cavity mode. In the optical
regime the application of cooling and trapping techniques in cavity
QED \cite{trapped-atoms} allows the deterministic loading of single
atoms in a high-finesse cavity, with accurate position control and
trapping times of many seconds \cite{Fortier}. In this regime
laser-assisted three-level atoms can behave as effective two-level
atoms with negligible spontaneous emission \cite{SDOAL}. On the
other hand, trapped atomic ions can remain in an optical cavity for
an indefinite time in a fixed position, where they can couple to a
single mode without coupling rate fluctuations \cite{ioncavity}.
These systems are quite promising to our purposes and could become
almost ideal in case of achievement of the strong coupling regime.\\
\indent Under full resonance conditions and starting from the vacuum
state of the cavity field, and for negligible atomic decays and
thermal fluctuations, we solved exactly the system dynamics for any
initial preparation of the atom pair, thus deriving a number of
analytical results on the whole system as well as the different
subsystems. These results are confirmed and extended by numerical
simulations, e.g. including also atomic
decays and investigating regimes under weaker driving conditions.\\
\indent Here we discuss some results mainly concerning the
bi-atomic subsystem. First of all we find that if atoms are prepared
in a separable state, no atom-atom entanglement can be generated, in
agreement with previous results that showed the classical nature of
atom-atom correlations under purely unitary dynamics~\cite{IJQI}. If
atoms are initially quantum correlated, their dynamics can be quite
different as it can be appreciated in the Bell basis.\\
\indent If the initial state is any linear combination of the states
$|\Phi^-\rangle$ and $|\Psi^-\rangle$, it does not evolve in time.
The atomic subspace is free from cavity dissipation and decoherence,
hence the initial entanglement can be preserved for storage times
useful
for quantum computation/communication purposes~\cite{Minns}.\\
\indent If the atoms are prepared in any superposition of the other
Bell states $|\Phi^+\rangle, |\Psi^+\rangle$ we find, remarkably,
that the same function describes the decay of atom-atom concurrence,
that is an entanglement measure, as well as of environment-induced
decoherence and purity. Also we show that joint atomic measurements
allow monitoring all these fundamental quantities. By generalizing
the one-atom analysis of Ref.~\cite{SDOAL} we can describe the
nontrivial decay process. First of all, in a short time/low
dissipation limit ($kt\ll1$) the whole system is in a pure,
entangled, cat-like state, as e.g. in Eq.~(\ref{evolvedrho})
starting from $|0\rangle\otimes|\Phi^+\rangle$. In this initial
stage the decay is quadratic in time and independent of the cavity
dissipative rate $k$. The subsequent behavior can be well
approximated by an exponential decay. In a weak coupling regime
($g/k\ll1$) the decoherence and disentanglement rate is $\gamma_D
\simeq 8g^2/k$. In a strong coupling regime ($g/k\gtrsim1$) the rate
is $\gamma_D \simeq 2g$, i.e., twice the JC coupling frequency. In
this regime the whole process is independent of the dissipative rate
$k$. It is in fact the strong coupling of each strongly driven atom
to the cavity field, which is in turn entangled with the
environment, that rules the decay of both entanglement and quantum
coherence of the bi-atomic subsystem.

\begin{acknowledgments}
E.S. thanks financial support from EuroSQIP and DFG SFB 631
projects, as well as from German Excellence Initiative via NIM.
\end{acknowledgments}

\appendix*
\section{Solution of the master equation}
We illustrate how to solve the master equation (\ref{eq:ME3}) in the
case of exact resonance, $\delta=0$.  For the density operator
$\hat{\rho}(t)$ of the whole system, we introduce the decomposition
of Eq.(\ref{decomposition}). Therefore, the ME is equivalent to the
following set of uncoupled equations for the field operators
$\hat{\rho}_{ij}(t)=\meanvalue{i}{\hat{\rho}(t)}{j}$:
\begin{eqnarray}\label{App:16operators}
&\dot{\hat{\rho}}_{11}&=-ig[\hat{a}^{\dag}+\hat{a},\hat{\rho}_{11}] + \hat{\mathcal{L}}_f\hat{\rho}_{11} \,\nonumber\\
&\dot{\hat{\rho}}_{12,13}&=-ig(\hat{a}^{\dag}+\hat{a})\hat{\rho}_{12,13} + \hat{\mathcal{L}}_f\hat{\rho}_{12,13} \,\nonumber\\
&\dot{\hat{\rho}}_{14}&=-ig\{\hat{a}^{\dag}+\hat{a},\hat{\rho}_{14}\} + \hat{\mathcal{L}}_f\hat{\rho}_{14}\,\nonumber\\
&\dot{\hat{\rho}}_{22,23,33}&=\hat{\mathcal{L}}_f\hat{\rho}_{22,23,33} \,\nonumber\\
&\dot{\hat{\rho}}_{24,34}&=-ig\hat{\rho}_{24,34}(\hat{a}^{\dag}+\hat{a})
+
\hat{\mathcal{L}}_f\hat{\rho}_{24,34} \,\nonumber\\
&\dot{\hat{\rho}}_{44}&=ig[\hat{a}^{\dag}+\hat{a},\hat{\rho}_{44}] +
\hat{\mathcal{L}}_f\hat{\rho}_{44}
\end{eqnarray}
where the brackets [ , ] and braces \{ , \} denote the standard
commutator and anti-commutator symbols and
$\dot{\hat{\rho}}_{j,i}(t)=[\dot{\hat{\rho}}_{i,j}(t)]^{\dagger}$.
In the phase space associated to the cavity field we introduce the
functions $\chi_{ij}(\beta,t)=
Tr_f[\hat{\rho}_{ij}(t)\hat{\mathcal{D}}(\beta)]$. We note that the
functions $\chi_{ij}(\beta,t)$ cannot be interpreted as
characteristic functions \cite{Glauber2} for the cavity field,
because the operators $\hat{\rho}_{ij}$ do not exhibit all required
properties of a density operator. As a consequence the functions
$\chi_{ij}(\beta,t)$ do not fulfill all conditions for quantum
characteristic functions. Nevertheless, they are continuous and
square-integrable, which is enough for our purposes. Equations
(\ref{App:16operators}) become in phase space
\begin{eqnarray}
\label{App:eq1} \dot{\chi}_{11}&=& ig(\beta + \beta^*)\chi_{11}\,\nonumber\\
&-&\frac{k}{2}\bigg( \beta\frac{\partial}{\partial
\beta}-\beta^*\frac{\partial}
{\partial \beta^*} +\vert\beta\vert ^2 \bigg ) \chi_{11} \,\nonumber\\
\dot{\chi}_{12,13}&=&-ig \left [ \frac{\partial}{\partial
\beta}-\frac{\partial}{\partial
\beta^*}-\frac{1}{2}(\beta+\beta^*)\right]
\chi_{12,13}\,\nonumber\\
&-&\frac{k}{2}\left ( \beta\frac{\partial}{\partial
\beta}-\beta^*\frac{\partial}{\partial \beta^*} +\vert\beta\vert
^2\right ) \chi_{12,13}\,\nonumber\\
\dot{\chi}_{14} &=&-2ig \left [ \frac{\partial}{\partial
\beta}-\frac{\partial}{\partial \beta^*}\right] \chi_{14}\,\nonumber\\
&-&\frac{k}{2}\left ( \beta\frac{\partial}{\partial
\beta}-\beta^*\frac{\partial}{\partial \beta^*} +\vert\beta\vert
^2\right ) \chi_{14}\,\nonumber\\
\dot{\chi}_{22,23,33}&=&-\frac{k}{2}\left (
\beta\frac{\partial}{\partial \beta}-\beta^*\frac{\partial}{\partial
\beta^*} +\vert\beta\vert
^2\right ) \chi_{22,23,33}\,\nonumber\\
\dot{\chi}_{24,34}&=&-ig \left [ \frac{\partial}{\partial
\beta}-\frac{\partial}{\partial
\beta^*}+\frac{1}{2}(\beta+\beta^*)\right]
\chi_{24,34}\,\nonumber\\
&-&\frac{k}{2}\left ( \beta\frac{\partial}{\partial
\beta}-\beta^*\frac{\partial}{\partial \beta^*} +\vert\beta\vert
^2\right ) \chi_{24,34}\,\nonumber\\
\dot{\chi}_{44}&=&-ig(\beta +
\beta^*)\chi_{11}\,\nonumber\\
&-&\frac{k}{2}\left( \beta\frac{\partial}{\partial
\beta}-\beta^*\frac{\partial}{\partial \beta^*} +\vert\beta\vert ^2
\right ) \chi_{44}.
\end{eqnarray}
For the initial atomic preparation  we consider the general pure
state $|\psi_{a}(0)\rangle=\sum_{i=1}^4 c_i |\nu_i\rangle$, where
the coefficients $c_i$ are normalized as $\sum_{i=1}^4|c_i|^2=1$ and
$\{|\nu_i\rangle \}_{i=1,..,4}=\{|\Phi^+\rangle, |\Phi^-\rangle,
|\Psi^+\rangle, |\Psi^-\rangle\}$ is the Bell basis. By this choice
we can describe atoms in a maximally or partially entangled state,
as well as separable states with both atoms in a superposition state
$\ket{\Psi}_{a}(0)=\big ( a_1\ket{e}_1+b_1\ket{g}_1\big )
\otimes\big ( a_2\ket{e}_2+b_2\ket{g}_2\big )$ where
\begin{eqnarray}
\label{App:pivsaj}
c_1=\frac{a_1a_2+b_1b_2}{\sqrt{2}}\hspace{0.5cm}c_2=\frac{a_1a_2-b_1b_2}{\sqrt{2}}\,\nonumber\\
c_3=\frac{a_1b_2+b_1a_2}{\sqrt{2}}\hspace{0.5cm}c_4=\frac{a_1b_2-b_1a_2}{\sqrt{2}}.
\end{eqnarray}
For the cavity field we consider the vacuum state, so that for the
whole system the initial state is
$\hat{\rho}(0)=\ket{0}\bra{0}\otimes|\psi_{a}(0)\rangle\langle
\psi_{a}(0)| $, corresponding to the functions
\begin{equation}
\label{App:chiijmagic} \chi_{ij}(\beta,0)=\left(\sum_{k,l=1}^4 c_k
c_l^*\langle i|\nu_k\rangle\langle\nu_l|j\rangle
\right)\exp\{|\beta|^2/2\}.
\end{equation}
Introducing the real and imaginary part of the variable $\beta$ we
can derive from Eq.(\ref{App:eq1}) a system of uncoupled equations
such that each of them can be solved by applying the method of
characteristics \cite{Barnett}. The solutions are:
\begin{eqnarray}
\label{App:solchibeta}
&\chi_{11,44}(\beta,t)&=\frac{1}{2}|c_1\pm c_3|^2\exp\left \{ -\frac{\vert\beta\vert^2}{2}
\mp\alpha^*(t)\beta\pm\alpha(t)\beta^*\right \}\,\nonumber \\
&\chi_{12,13}(\beta,t)&=-\frac{1}{2}(c_1+c_3)(c_2\mp
c_4)^*f_1(t)\times\,\nonumber \\
&&\exp\left \{ -\frac{\vert\beta\vert^2}{2}+\alpha(t)\beta^*\right
\}
\,\nonumber\\
&\chi_{14}(\beta,t)&=\frac{1}{2}(c_1+c_3)(c_1-c_3)^*f_2(t)\times\,\nonumber \\
&&\exp\left \{ -\frac{\vert\beta\vert^2}{2}+\alpha(t)^*\beta+\alpha(t)\beta^*\right \} \,\nonumber \\
&\chi_{22,33}(\beta,t)&=\frac{1}{2}|c_2\mp c_4|^2\exp \left \{ -\frac{\vert\beta\vert^2}{2} \right \} \,\nonumber \\
&\chi_{23}(\beta,t)&=\frac{1}{2}(c_2-c_4)(c_2+c_4)^*\exp \left \{ -\frac{\vert\beta\vert^2}{2} \right \} \,\nonumber \\
&\chi_{24,34}(\beta,t)&=-\frac{1}{2}(c_2\mp
c_4)(c_1-c_3)^*f_1(t)\times\,\nonumber \\
&& \exp\left \{
-\frac{\vert\beta\vert^2}{2}+\alpha^*(t)\beta\right \}
\end{eqnarray}
where $\alpha(t)$ and $f_1(t)$ are defined in Eqs.(\ref{alfat}),
(\ref{eq:f1}) and $f_2(t)=f_1^4(t)$. From the above expressions we
can recognize that the corresponding field operators
$\hat{\rho}_{ij}(t)$ are:
\begin{eqnarray}
\label{App:rhoijBell}
&\hat{\rho}_{11,44}(t)&=\frac{1}{2}|c_1\pm c_3|^2\ketbra{\mp\alpha(t)}{\mp\alpha(t)} \,\nonumber \\
&\hat{\rho}_{12,13}(t)&=\frac{1}{2}(c_1+c_3)(c_2\mp c_4)^*\frac{f_1(t)}{e^{-\frac{|\alpha(t)|^2}{2}}}\ketbra{-\alpha(t)}{0} \,\nonumber \\
&\hat{\rho}_{14}(t)&=\frac{1}{2}(c_1+c_3)(c_1-c_3)^*\frac{f_2(t)}{e^{-2|\alpha(t)|^2}}\ketbra{-\alpha(t)}{\alpha(t)} \,\nonumber \\
&\hat{\rho}_{22,33}(t)&=\frac{1}{2}|c_2\mp c_4|^2\ketbra{0}{0} \,\nonumber \\
&\hat{\rho}_{23}(t)&=\frac{1}{2}(c_2-c_4)(c_2+c_4)^*\ketbra{0}{0} \,\nonumber \\
&\hat{\rho}_{24,34}(t)&=-\frac{1}{2}(c_2\mp c_4)(c_1-c_3)^*\frac{f_1(t)}{e^{-\frac{|\alpha(t)|^2}{2}}}\ketbra{0}{\alpha(t)}. \,\nonumber \\
\end{eqnarray}
Hence, by remembering that
$\hat{\rho}_{ji}=\hat{\rho}_{ij}^{\dagger}$, we can reconstruct the
whole system density operator $\hat{\rho}(t)$ of Eq.
(\ref{decomposition}).

\end{document}